\documentclass[a4paper,11pt]{article}
\pdfoutput=1
\usepackage{jheppub}
\usepackage{amsmath,amssymb,euscript}
\usepackage{slashed}
\usepackage{xspace}
\usepackage{color}
\usepackage{accents}
\usepackage{hyperref}
\usepackage{epsfig}
\usepackage{xcolor}
\usepackage{xspace}
\usepackage{verbatim}
\usepackage{multirow}
\usepackage{booktabs,graphicx}
\usepackage{mathtools}
\usepackage{tabulary}

\newcolumntype{K}[1]{>{\centering\arraybackslash}p{#1}}

\emergencystretch=\maxdimen
\newcommand{\tev}{\, \mathrm{TeV}}
\newcommand{\gev}{\, \mathrm{GeV}}

\newcommand{\dzero}{d_{0}}
\newcommand{\bbbar}{b\bar{b}}
\newcommand{\met}{\slashed{p}_T}

\newcommand{\nn}{\mathrm{NN}}
\newcommand{\ctauc}{c \tau_c}

\begin{document}
\title{Soft displaced leptons at the LHC}

\author[1]{Freya Blekman,}
\author[2]{Nishita Desai,}
\author[3]{Anastasiia Filimonova,}
\author[1]{Abanti Ranadhir Sahasransu,}
\author[3]{Susanne Westhoff\,}

\affiliation[1]{Inter-university Institute for High Energies, Vrije Universiteit Brussel, 1050 Brussels, Belgium}
\affiliation[2]{Department of Theoretical Physics, Tata Institute of Fundamental Research, Mumbai 400005, India}
\affiliation[3]{Institute for Theoretical Physics, Heidelberg University, 69120 Heidelberg, Germany}

\emailAdd{freya.blekman@vub.be}
\emailAdd{nishita.desai@tifr.res.in}
\emailAdd{filimonova@thphys.uni-heidelberg.de}
\emailAdd{abanti.ranadhir.sahasransu@vub.be}
\emailAdd{westhoff@thphys.uni-heidelberg.de}

		\begin{flushright}
		TIFR/TH/20-21\\
		P3H-20-029\\
		\end{flushright}

\abstract{Soft displaced leptons are representative collider signatures of compressed dark sectors with feeble couplings to the standard model. Prime targets are dark matter scenarios where co-scattering or co-annihilation sets the relic abundance upon freeze-out. At the LHC, searches for soft displaced leptons are challenged by a large background from hadron or tau lepton decays. In this article, we present an analysis tailored for displaced leptons with a low transverse momentum threshold at $20\gev$. Using a neural network, we perform a comprehensive analysis of the event kinematics, including a study of the expected detection efficiencies and backgrounds at small momenta. Our results show that weak-scale particles decaying into soft leptons with decay lengths between $1\,\rm mm$ and $1\,\rm m$ can be probed with LHC Run 2 data. This motivates the need for dedicated triggers that maximize the sensitivity to displaced soft leptons.}

\maketitle
\flushbottom

\clearpage
%---------------------------------------------------------------------------
\section{Introduction}
\label{sec:intro}
\noindent Collider searches for long-lived particles open a new dimension in the hunt for new physics. Lifetime measurements are particularly sensitive to hidden sectors of new particles with tiny couplings to the Standard Model (SM), well below the strength of weak interactions~\cite{Alimena:2019zri,Beacham:2019nyx}. If one of the states from the hidden sector can be produced at a sizeable rate, its decay is suppressed by the tiny coupling to the hidden sector, and visible decay products appear displaced from the collision point. Such signatures are naturally predicted in dark matter scenarios beyond the thermal WIMP (for an overview see Ref.~\cite{Battaglieri:2017aum}). The potential to discover feebly coupling dark matter through long-lived particles at colliders is unique, as potential signals in direct and indirect detection experiments are often suppressed by the tiny interaction.

Tiny couplings not only affect the collider phenomenology, but also the thermal history of dark matter. In scenarios where the relic abundance is produced via thermal freeze-out, pair annihilation is suppressed and no longer efficient around the freeze-out temperature. Instead, the relic abundance is set by processes of co-annihilation~\cite{Griest:1990kh} or co-scattering~\cite{DAgnolo:2017dbv,Garny:2017rxs} with another particle from the dark sector, followed by efficient annihilation of this particle. In what follows we will call this extra state a \emph{dark partner}. Freeze-out through co-annihilation or co-scattering can be naturally realized for dark partners $\chi'$ with small couplings to dark matter $\chi$ and sizeable couplings to SM particles. In addition, the number densities of dark matter and its partner around the freeze-out temperature need to be comparable, in order to ensure efficient co-annihilation or co-scattering~\cite{Griest:1990kh}. This results in an upper bound on the mass difference. These conditions set a target for collider searches: small dark matter couplings lead to {\it displaced} decays $\chi'\to \chi f$ with visible decay products $f$. If in addition the mass spectrum is compressed, as suggested by the requirement of abundant partners at freeze-out, the decay products are {\it soft}. In many models with weak-scale dark states the transverse momenta of the visible decay products range around $10-40\gev$~(see for example Ref.~\cite{Baker:2015qna}). The decay length of the dark partner can vary between zero and a few centimeters for co-annihilation and up to a meter for co-scattering~\cite{Garny:2017rxs,DAgnolo:2017dbv,Garny:2018icg, Bharucha:2018pfu,Filimonova:2018qdc,Junius:2019dci,DAgnolo:2019zkf}, depending on the model. For even smaller dark matter couplings, dark matter is never in thermal equilibrium and the relic abundance can be produced through freeze-in~\cite{Hall:2009bx} or the decay of thermal dark states ~\cite{McDonald:2001vt}, which results in even larger decay lengths at colliders~\cite{Shoemaker:2010fg,Evans:2016zau,Belanger:2018ccd,Junius:2019dci}.

At the LHC, searches for soft displaced objects face substantial experimental challenges. Signals with small momenta {\it and} small displacements are produced with a large background from displaced decays of $B$ hadrons, commonly referred to as {\it heavy flavor} (HF). In current searches, this background is rejected by trigger settings or other selection criteria, which means that these searches also reject signals with small transverse momenta. The LHC collaborations are currently designing new trigger menus for LHC Run 3. In this work, we examine strategies to reject the background in searches for signals with soft displaced leptons. Although trigger selections are subject to different requirements than offline analysis, the results can provide input for future trigger strategies.

Despite the challenges, existing searches for soft {\it or} displaced particles have been performed that probe co-annihilating or co-scattering dark matter in parts of the predicted signal space. Searches for prompt soft leptons by ATLAS~\cite{Aaboud:2017leg} and CMS~\cite{Sirunyan:2018iwl} trigger on events with a high-energetic jet from the initial state. They probe co-annihilation in scenarios with promptly decaying dark partners, for instance, supersymmetric electroweakinos. In scenarios with charged states, lifetimes of a few nanoseconds can be probed by searches for disappearing charged tracks~\cite{Aaboud:2017mpt,Sirunyan:2018ldc,Sirunyan:2020pjd}. Due to the detector properties, these searches are only sensitive to long decay lengths and require that the charged decay products escape the detector. The applicability of disappearing track searches to dark sectors is, therefore, model-dependent and limited to this specific region of the signal space.

Two searches for displaced leptons have been performed by CMS, one at $8\tev$~\cite{Khachatryan:2014mea} and an updated search at $13\tev$~\cite{CMS:2016isf}. However, the $8\tev$ analysis is not sensitive to compressed dark sectors with electroweak production rates due to the limited data set. The $13\tev$ analysis uses a lepton trigger with a hard momentum threshold of $40\gev$, which drastically reduces the sensitivity to dark sectors with small mass splittings. Both searches are sensitive to {\it hard} displaced leptons from dark partners with decay lengths between about $1\,$cm and $1\,$m.

In this work, we perform a first analysis making use of soft displaced leptons at the $13\tev$ LHC. We show that a search for leptons with $p_T > 20\gev$ can probe weak-scale particles with decay lengths from $1\,$mm to $2\,$m and a mass splitting with the lightest dark state of around $20\gev$. This search is optimal to probe scenarios of co-annihilating and co-scattering dark matter by covering the previously unexplored phase space region.

This article is organized as follows. In Sec.~\ref{sec:theory}, we discuss the phenomenology of co-annihilation and co-scattering dark matter and define benchmark scenarios for soft displaced leptons at the LHC. In Sec.~\ref{sec:lhcsignal}, we analyze the phenomenology of these signal benchmarks and perform a detailed analysis of the expected background from heavy flavor decays at low transverse momenta. In Sec.~\ref{sec:mva}, we describe our multivariate analysis and show the gain in performance compared to a simple cut-based analysis. Our predictions for the LHC based on Run 2 and Run 3 data are presented in Sec.~\ref{sec:predictions}. By comparing the sensitivity of our proposed analysis with existing searches, we demonstrate that a dedicated search for soft displaced leptons covers the entire phase-space region predicted by compressed hidden sectors that is currently still unexplored. We conclude in Sec.~\ref{sec:conclusions} with suggestions for the experimental realization of future searches for soft displaced leptons.

%---------------------------------------------------------------------------
\section{Soft displaced leptons from dark sectors}\label{sec:theory}
\noindent At the LHC, signals with soft particles and missing energy are often predicted in scenarios with compressed dark sectors around the weak scale. A minimal version of a compressed dark sector consists of a stable neutral dark matter candidate $\chi^0$ with mass $m_0$ and a charged dark partner $\chi^+$ with mass $m_c$ and a small mass splitting

\begin{align}
\Delta m = m_c - m_0 \ll m_0\,.
\end{align}

If the dark partner carries electroweak charge, its production at the LHC proceeds through weak interactions and leads to signatures like
\begin{align}\label{eq:lhc-prod}
pp\to \chi^+\chi^- \to (\chi^0 f)(\chi^0 f)\,.
\end{align}
Due to the small mass splitting $\Delta m$ among the dark states, the standard-model decay products $f$ carry little momentum, provided that the dark partner is produced at moderate boost. Depending on the decay, $f$ can be one or several particles, at least one of them carrying electric charge. A well known example of such a process are supersymmetric charginos decaying into neutralinos and leptons or jets~\cite{Djouadi:2001fa,ArkaniHamed:2006mb,Bramante:2015una,Canepa:2020ntc}.

\subsection{Co-annihilating and co-scattering dark matter}
Scenarios of feebly interacting dark matter, where the freeze-out in the early universe is set by co-annihilation or co-scattering, often predict a compressed spectrum. We focus on dark states around the weak scale, which are usually non-relativistic at freeze-out, so that their number densities $n_0$ and $n_c$ scale exponentially with the freeze-out temperature $T_f$. As a consequence, the {\it relative} number density of $\chi^\pm$ and $\chi^0$ scales exponentially with the mass difference as
\begin{align}\label{eq:number-densities}
\frac{n_c}{n_0} \sim \exp(-\Delta m/T_f)\,.
\end{align}
For efficient dark matter interactions with the charged partner around the freeze-out temperature $T_f\approx 0.2\,m_0$, both dark states need to be abundant. This results in a (model-dependent) upper bound on the mass difference.\footnote{This upper bound on the mass difference can be relaxed if the dark partner freezes out before dark matter~\cite{DAgnolo:2017dbv}. In this case, the relic abundance is typically set by the rate of partner annihilations.} In many models with weak-scale dark states, freeze-out through co-annihilation and co-scattering predicts a mass difference of about
\begin{align}\label{eq:deltam}
   10\gev \lesssim \Delta m \lesssim 40\gev\,.
\end{align}
Which processes determine the relic abundance upon freeze-out mostly depends on the coupling of dark matter to the dark partner. For sizeable couplings pair annihilation $\chi^0\chi^0 \to ff$ is efficient around the freeze-out temperature and sets the relic abundance. When successively decreasing the coupling, pair annihilation becomes inefficient and co-annihilation processes $\chi^0\chi^+ \to ff$ set the relic abundance instead. For even smaller couplings, co-annihilations become inefficient at freeze-out as well. The relic abundance is now set by co-scattering of dark matter with particles $f$ from the thermal bath, followed by efficient annihilation of the heavier dark partners
\begin{align}
\chi^0 f \to \chi^+ f' \quad \Longrightarrow \quad \chi^+\chi^- \to f f\,.
\end{align}
For weak-scale dark sectors, freeze-out via co-annihilation or co-scattering typically occurs at dark matter couplings to bath particles in the range
\begin{align}\label{eq:coupling}
10^{-3} > g_\chi > 10^{-8}\,.
\end{align}
At even smaller couplings, dark matter leaves chemical or even kinetic equilibrium well before freeze-out, and a different mechanism has to be invoked to explain the observed relic abundance.

%%%%%%%%%%%%%%%%%%%%%%%%%%%%%%%%%%%%%%%%%%%%%%%%%%%%%%%
\subsection{Signal characteristics}\label{sec:signal}
\begin{figure}[!t]
	\centering
	\includegraphics[width=.37\textwidth]{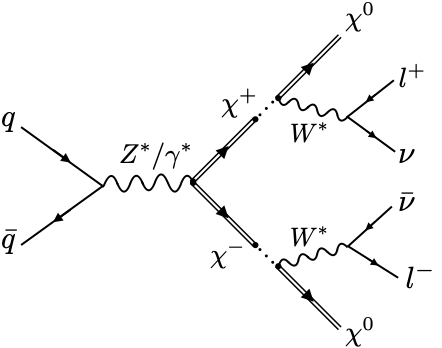}
	\caption{Feynman diagram for a signal with soft displaced leptons from decays of dark states at hadron colliders. 
	\label{fig:feynman}}
\end{figure}

\begin{figure}[tp]
\centering
	\includegraphics[width=.6\textwidth]{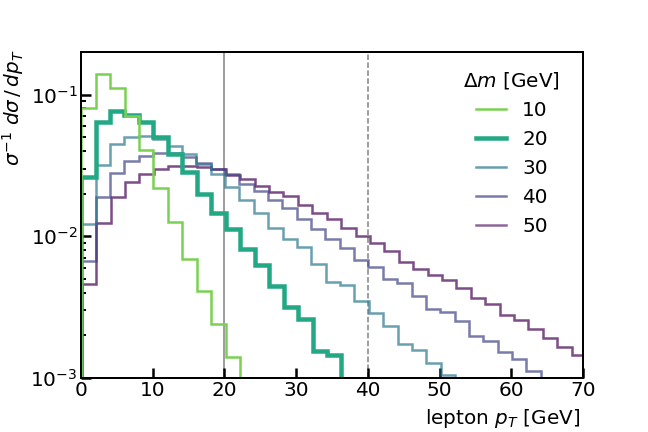}
	\hspace*{0.5cm}
\caption{Transverse momentum distribution of the lepton $\ell^+$ in $pp\to \chi^+\chi^- \to (\chi^0\ell^+\nu)(\chi^0\ell^-\bar{\nu})$ at $\sqrt{s} = 13\tev$ for $m_c = 324\gev$ and various mass splittings $\Delta m$, normalized to the total cross section. The distribution at $\Delta m = 20\gev$ (thick green) is typical for weak-scale dark matter. The dashed line shows the lepton momentum threshold in~\cite{CMS:2014hka,CMS:2016isf}; the solid line indicates the threshold in our analysis. Based on event generation at parton level using {\tt MadGraph5\_aMC@NLO}.}
\label{fig:pt-leptons}
\end{figure}

\noindent Remarkably, the described scenarios of co-annihilating and co-scattering dark matter predict signatures with soft displaced particles within the acceptance of the LHC detectors. Our analysis focuses on final states with soft displaced leptons and missing energy, produced from dark states with electroweak couplings via
\begin{align}\label{eq:signal-process}
    pp\to Z^\ast / \gamma^\ast \to \chi^+\chi^- \to (\chi^0\ell^+\nu)(\chi^0\ell^-\bar{\nu})\,.
\end{align}
In Fig.~\ref{fig:feynman} we show the corresponding Feynman diagram.
%-------------------------------------------------------------------------
%-------------------------------------------------------------------------
 Due to the small mass splitting, the dark partners decay via an off-shell $W$ boson. The rate and kinematics of this signature are fully determined by the three variables
\begin{align}\label{eq:variables}
(m_c,\  \Delta m,\  \ctauc )\,,
\end{align}
where $\ctauc$ is the nominal decay length of the dark partner. The partner's mass $m_c$ determines the $\chi^+\chi^-$ production rate. The mass splitting $\Delta m$ is set by requiring the observed relic abundance. In Fig.~\ref{fig:pt-leptons} we show the transverse momentum distribution of the lepton $\ell^+$ for $m_c = 324\gev$ and various $\Delta m$, without applying any kinematic cuts.
%-------------------------------------------------------------------------
%-------------------------------------------------------------------------
For $\Delta m = 20\gev$, as predicted by co-annihilation or co-scattering at this mass scale, the distribution peaks at $p_T(\ell) \approx 7\gev$. Therefore the ballpark of events fall well below the threshold of conventional lepton triggers used by ATLAS and CMS~\cite{Sirunyan:2020zal, Aad:2020uyd,Aad:2019wsl}, as indicated by the dashed line. The main goal of our analysis is to lower the threshold to $p_T > 20\gev$ (black line) to be sensitive to these dark matter scenarios. Notice that the mass scale of the dark partner has little impact on the spectrum at low momenta. In the relevant mass range $m_c=100-500\gev$, it only mildly affects the tail of the distribution.

The nominal decay length $\ctauc$ strongly depends on the dark matter coupling $g_\chi$, see Eq.~\eqref{eq:coupling}. In models with weak interactions, the three-body partial decay width can be estimated for $\Delta m \ll m_0 \sim m_W$ as~\cite{Djouadi:2001fa}
\begin{align}\label{eq:gamma-3body}
\Gamma_c \approx \frac{g_\chi^2 G_F}{30 \sqrt{2} \pi^3}\frac{(\Delta m)^5}{m_W^2} = 1.6\times 10^{-14}\,\gev \left(\frac{g_\chi}{6\times10^{-5}}\right)^2\left(\frac{\Delta m}{20\,\gev}\right)^5\,,
\end{align}
where we have chosen typical parameters for weak-scale co-scattering as a reference. Assuming that no other decay channels are accessible, these reference parameters correspond to a proper lifetime and nominal decay length of the 
dark partner around

\begin{align}\label{eq:ctau-coscattering}
\tau_c = \frac{1}{\Gamma_c} \approx 41\,\text{ps}\,,\qquad \ctauc \approx 1\,\text{cm}\,.
\end{align}
More generally, weak-scale co-scattering scenarios can predict decay lengths up to a few meters. For co-annihilation the decay length is typically shorter, due to the larger dark matter coupling. The nominal decay length is related to the decay length in the lab frame as
\begin{align}\label{eq:decay-length}
    d = (\beta\gamma)\,\ctauc\,,
\end{align}
where $(\beta\gamma)$ is the Lorentz boost of the dark partner. Due to the exponential decay probability, the number of dark partners decaying within a sphere of radius $d$ around the production point is given by
\begin{align}\label{eq:exp-decay}
    N(d) = N(0)\int_0^d \frac{dr}{d} \exp\left(-\frac{r}{d}\right),
\end{align}
where $N(0)$ is the number of dark states produced at the collision point. The decay length $d$ is not directly observable at the LHC. To describe the displacement of the leptons, we employ the widely used unsigned transverse impact parameter $\dzero$, defined as the distance of closest approach of the lepton track from the collision point in the azimuthal plane. Following Ref.~\cite{CMS:2016isf}, identification of displaced leptons requires that the transverse impact parameter should lie within the range
 \begin{align}\label{eq:impact-parameter}
     200\,\mu\text{m} < \dzero < 10\,\text{cm}\,.
 \end{align}
 \noindent The range of nominal decay lengths $\ctauc$ that can be probed within this range of $\dzero$ depends on the overall boost of the dark states, which determines the decay length $d$ (see Eq.~\eqref{eq:decay-length}), and on the transverse component of the boost. For $m_c < 500\gev$, most particles are produced with boosts within $0.2 < (\beta\gamma) < 5$, with a peak around $(\beta\gamma) \approx 1$. Highly boosted, {\it i.e.}, light dark states tend to be emitted along the beam line, leading to a smaller transverse decay length $\dzero < d$. In a typical scenario with $m_c = 220\gev$ and $\Delta m = 20\gev$, the dark partner has a transverse momentum of $p_T(\chi^\pm) \approx 100 \gev$ and the decay lepton carries $p_T(\ell) \approx 10\gev$. In this case the observable range of $\dzero$ from Eq.~\eqref{eq:impact-parameter} roughly corresponds to
 \begin{align}\label{eq:ctau-range}
 1\,\text{mm} \lesssim \ctauc \lesssim 20\,\text{cm}\,.
 \end{align}
By comparing with Eq.~\eqref{eq:ctau-coscattering}, we see that the LHC detectors are well suited to probe co-scattering dark matter, as well as co-annihilation with sufficiently small dark matter couplings.

%-------------------------------------------------
\subsection{Benchmarks}\label{sec:benchmarks}
\noindent Depending on the underlying model, the production cross section, lifetime and decay channels of the dark states may vary. For concreteness, we consider a specific model for co-scattering dark matter in which we can predict the three characteristic variables $(m_c,\Delta m, \ctauc)$ in terms of fundamental parameters. The dark sector consists of two fermion fields, transforming under weak interactions as a singlet and an adjoint triplet with vector-like couplings. Upon electroweak symmetry breaking the neutral components of these fields mix through an effective scalar interaction with the Higgs field. The mixing $\theta$ leads to a dark sector with three mass eigenstates $\chi_\ell^0,\chi^\pm,\chi_h^0$. The lightest state  $\chi_\ell^0 = \chi^0$ is a stable dark matter candidate and the charged states $\chi^\pm$ act as dark partners. Co-scattering is realized for $(m_c - m_0)/m_0 \approx 0.1$ and small $\theta$.
 For details we refer the reader to Ref.~\cite{Filimonova:2018qdc}. The setup is similar to the bino-wino scenario in supersymmetry with decoupled higgsinos~\cite{Nagata:2015pra}. 

Motivated by this model, we define specific scenarios with soft displaced lepton signals as benchmarks for our analysis. In Tab.~\ref{tab:benchmarks}, benchmarks 1 and 2 correspond to dark matter scenarios, where $\Delta m$ and $\theta$ are chosen such that the observed relic abundance is obtained for dark partner lifetimes in reach of the LHC, see Eq.~\eqref{eq:ctau-range}.
%-----------------------------------------------------------
\begin{table}[tp]\begin{center}
\renewcommand{\arraystretch}{1.2}
\setlength{\tabcolsep}{4mm}
\begin{small}\begin{tabular}{c|c|c|c|c}
\toprule
\text{\#} & $m_c$ [GeV] & $\Delta m$ [GeV] & $\ctauc$ [cm] & $\mathcal{B}(\ell^+ \ell^-)$\\\midrule[1pt]
1 & 324 & 20 & 2 & 0.025 \\ 
2 & 220 & 20 & 3 & 0.014 \\
\midrule[0.5pt]
3 & 220 & 20 & 0.1 & 1 \\
4 & 220 & 20 & 1 & 1 \\
5 & 220 & 20 & 10 & 1 \\
6 & 220 & 20 & 100 & 1 \\
\midrule[0.5pt]
7 & 220 & 40 & 1 & 1 \\
\bottomrule[1pt]
\end{tabular}\end{small}
\end{center}
\caption{\label{tab:benchmarks}Benchmarks for soft displaced leptons $\ell = e,\,\mu$ at the LHC. Benchmarks 1 and 2 correspond to dark matter scenarios; benchmarks 3 to 6 feature soft leptons from dark partners with different decay lengths; in benchmark 7 we choose a larger mass splitting $\Delta m$ for comparison with previous analyses. In the last column we show the total branching ratio into lepton pairs. For benchmarks 1 and 2 see Eq.~\eqref{eq:br}.}
\end{table}
%----------------------------------------------------

In our dark matter model the dark partner decays via weak interactions, leading to final states with leptons and hadrons with branching ratios determined by their gauge quantum numbers. The branching ratio into a specific lepton flavor $\ell$ can be estimated as
\begin{equation}\label{eq:br}
    \mathcal{B}(\chi^+\to \chi^0\ell\,\nu)\sim \frac{\Gamma_\ell}{3 \Gamma_\ell + 2 N_c \Gamma_\ell +\Gamma_\pi}\,.
\end{equation}
Here $\Gamma_\ell$ and $\Gamma_\pi$ are the partial decay widths into leptons and pions, $N_c = 3$ is the number of colors and CKM mixing is approximated by $V_{ud} = V_{cs} = 1$. In our analysis we use numerical predictions for the three-body partial decay rates, see Sec.~\ref{sec:signal}. Two-body decays can play a role if $\chi^+$ is part of a larger $SU(2)$ multiplet. In our model, the decay $\chi^+ \to \chi_h^0\,\pi^+$ can open up for small mixing $\theta \lesssim 10^{-5}$, due to electroweak corrections~\cite{Ibe:2012sx}. The partial decay rate is~\cite{Filimonova:2018qdc}
\begin{align}\label{eq:gamma-pi}
\Gamma_\pi = \frac{2G_F^2}{\pi} |V_{ud}|^2 f_\pi^2 \cos^2\theta\, |\Delta m_{hc}|^3\Big(1 - \frac{m_\pi^2}{(\Delta m_{hc})^2}\Big)^{1/2}\,,
\end{align}
where $\Delta m_{hc}=m_h-m_c \gtrsim 140\,\text{MeV}$ is the mass difference between $\chi_h^0$ and $\chi^\pm$, and $f_\pi \simeq 130\,\text{MeV}$ is the pion decay constant. Decays into pions are relevant in benchmark 2.

 To explore the full discovery potential of the LHC, we extend the scope of our analysis beyond this model and cover a broader range of displaced soft lepton signals. To this end, we fix the dark partner's  mass to $m_c = 220\gev$ and vary the lifetime (benchmarks~3 to 6) and the mass splitting (benchmark~7). Smaller masses can be probed down to the LEP bound of $m_c \gtrsim 100\gev$. For larger masses the sensitivity at the LHC is statistically limited as the cross section rapidly decreases.
 
 In benchmarks~3 to 7 we assume that the dark partner  decays exclusively into electrons and muons. Such an assumption is realistic in many scenarios with hidden sectors that directly couple to leptons. Examples are feebly coupled leptophilic dark matter~\cite{Fox:2008kb,Freitas:2014jla,Evans:2016zau,Barducci:2018esg,Junius:2019dci} or models with heavy neutral leptons~\cite{Cai:2017mow,Drewes:2019fou}. This assumption maximizes the expected signal rate. It also makes the results of our analysis less model-dependent and easier to reinterpret in other scenarios.
 
 Soft displaced light leptons do not necessarily rely on a compressed hidden sector, but can also arise in decay chains involving leptonic tau decays. Such signatures are predicted for instance in models with pair-produced supersymmetric staus~\cite{Evans:2016zau}. While our analysis targets compressed spectra, its main features apply as well for soft displaced leptons as products of decay chains.
 
%---------------------------------------------------------------------------
\section{LHC signals of soft displaced leptons}\label{sec:lhcsignal}
Three kinds of searches are potentially sensitive to models with co-annihilation or co-scattering described above, {\it i.e.}, searches for prompt soft leptons from compressed spectra, searches for displaced leptons, and searches for disappearing tracks. In searches for prompt soft leptons ~\cite{Aaboud:2017leg,Sirunyan:2018iwl}, a high-energetic jet from initial-state radiation was used for triggering, as well as to enhance the amount of missing energy and the boost of the visible final-state particles~\cite{Schwaller:2013baa}. Such requirements, however, come at the cost of reducing the signal rate, which in the context of dark matter decreases the search sensitivity for heavy dark states. The current limit for promptly decaying weak triplet states  with a mass splitting of $\Delta m = 20\gev$ stands at $220\gev $~\cite{Bharucha:2018pfu, Filimonova:2018qdc}. As both searches require leptons with transverse impact parameters $\dzero \lesssim 0.1\,\text{mm}$, they are largely insensitive to long-lived dark states.

Searches for disappearing tracks probe the other end of the lifetime range~\cite{Aaboud:2017mpt,Sirunyan:2018ldc,Sirunyan:2020pjd}. The latest CMS search excludes charged particles with decay lengths larger than a few centimeters~\cite{Sirunyan:2020pjd}, while ATLAS is sensitive only to much larger $c\tau$ because of the different detector geometry. However, these bounds only apply if the decay products of the original charged particle are invisible in the detector. The presence of an extra lepton in the final state may degrade the sensitivity, either by causing the original track to fail the isolation criteria, or because the kink from the lepton track modifies the reconstruction of the original track from the charged state. This reduces the sensitivity to dark matter scenarios predicting soft displaced particles in the final state.

The sensitivity gap between searches for prompt soft leptons and disappearing charged tracks corresponds to decay lengths $c\tau_c$ from about $0.1$ mm to several centimeters. In this region, any leptons produced in decays of dark states are not identified as prompt leptons, and the track of the charged state is not long enough to be classified even as a ``disappearing" track. With our analysis we cover this gap by looking for decay products that are displaced, {\it i.e.}, have a transverse impact parameter larger than the detector resolution $d_0\gtrsim 200\,\mu\text{m}$.

In this paper, we look at events with one electron and one muon, both with large transverse impact parameters.  We choose to restrict to $e-\mu$  because we need a reliable data-driven estimate of backgrounds and this is the only published data currently available.  Further systematic backgrounds are expected for the  $\mu-\mu$ final state that cannot be estimated reliably by extrapolating the $e-\mu$ backgrounds.  In no particular order, these would include radiation from the experimental cavern, muons created by LHC beam interactions outside the experimental interaction region, cosmic rays, and muons originating from punch-through of hadronic jets.  The sensitivity of the $e-e$ final state is expected to be lower than the $e-\mu$ since electron identification efficiency degrades much more rapidly than muons with increasing displacement.  We therefore will restrict this study to the $e-\mu$ final state only.  However, both $e-e$ and $\mu-\mu$ final states should be explored in the future as important flavour non-universal effects may be missed by restricting to only one final state.

%%%%%%%%%%%%%%%%%%%%%%%%%%%%%%%%%%%%%%%%%%%%%%%%%%%%%%%%%%%%%%
\subsection{Existing displaced lepton searches}\label{sec:previous}
\noindent Several searches for displaced leptons have already been performed at the LHC~\cite{Khachatryan:2014mea, CMS:2014hka,CMS:2016isf,Aad:2019tcc,Aad:2019kiz}. Of these, the analyses in Refs.~\cite{CMS:2014hka, Aad:2019tcc,Aad:2019kiz} rely on reconstructing two leptons from the {\it same} vertex; therefore they are not sensitive to signals with two single leptons as in our scenarios.

We design our search for two individual displaced leptons without any vertex requirement, which covers a broader class of event topologies. Refs.~\cite{Khachatryan:2014mea, CMS:2016isf} describe searches for such individual displaced leptons from event topologies close to our dark matter scenarios, see Eq.~\eqref{eq:signal-process}.  The searches at 8 TeV~\cite{Khachatryan:2014mea} and 13 TeV~\cite{CMS:2016isf} differ mainly in the requirement of the lepton transverse momenta and therefore will be described together.  

Both analyses have been performed in a largely model-independent way, in particular without imposing many kinematic restrictions on the leptons or requiring further activity in the event. They select events with one electron and one muon of opposite charge, produced inside the active detection region of the CMS experiment with pseudorapidity $|\eta(\ell)|< 2.4$, $\ell = e,\,\mu$. At 13 TeV the transverse momentum requirements are $p_T(e) > 42 \gev$ and $p_T(\mu) > 40 \gev$; at 8 TeV both leptons satisfy $p_T(\ell) > 25\gev$. In the 13 TeV incarnation, the strong selection requirement on the lepton transverse momenta is necessary due to the trigger used for the analysis. 

The displacement of the leptons is observable through the unsigned impact parameter $\dzero$, defined as the distance of closest approach in the azimuthal plane of the lepton track to the collision point. Based on the impact parameter, three exclusive signal regions (SR) are defined as~\footnote{In the $8\tev$ analysis~\cite{Khachatryan:2014mea}, SR III is bounded by $d_0 < 2\,\text{cm}$.}
\begin{align}
\label{eq:sr}
\textrm{SR III}:\quad & \textrm{both leptons satisfying}~1\,\mathrm{mm} < \dzero < 10\,\mathrm{cm}
\nonumber \\
\textrm{SR II\ }: \quad & \textrm{at least one lepton failing SR III and both with}~\dzero > 500\,\mu\mathrm{m}\\
\textrm{SR I\ \ }: \quad &\textrm{at least one lepton failing SR II and both with}~\dzero > 200\,\mu\mathrm{m}\,. \nonumber 
\end{align}
We will adopt these signal regions for our analysis.

Since neither search observed an excess of events over the expected background, 95\% C.L. upper limits on the signal were obtained. To compare our results with the previous analyses, we use the Poisson log likelihood ratio
\begin{equation}
    \label{eqn:LLR}
    \mathcal{Q} = \sum_{i=\text{I,II,III}} - 2\log\left(\frac{\mathcal{L}_{S_i+B_i}}{\mathcal{L}_{B_i}}\right),\qquad \mathcal{L}_{S_i+B_i} = e^{-(S_i + B_i)}\frac{(S_i + B_i)^{N_i}}{N_i!}\,,
\end{equation}
where $N_i$ is the observed number of events in signal region $i$, $S_i$ and $B_i$ are the expected signal and background events in region $i$, and $N = \sum_i N_i$ is the total number of observed events. The 95\% C.L. upper limit for one measurement in 3 signal regions corresponds to $\mathcal{Q} = 5.99$. We define the ratio 
\begin{align}
    \label{eqn:R95}
    R_{95} = \mathcal{Q}/5.99\,,
\end{align}
so that $R_{95} = 1$ corresponds to the exclusion limit at the $95\%$ C.L. We checked explicitly that the likelihood ratio $\mathcal{Q}$ reproduces the exclusion limits from Ref.~\cite{Khachatryan:2014mea} within 1$\sigma$.

\begin{figure}[t]
	\centering
	\includegraphics[width=0.75\textwidth]{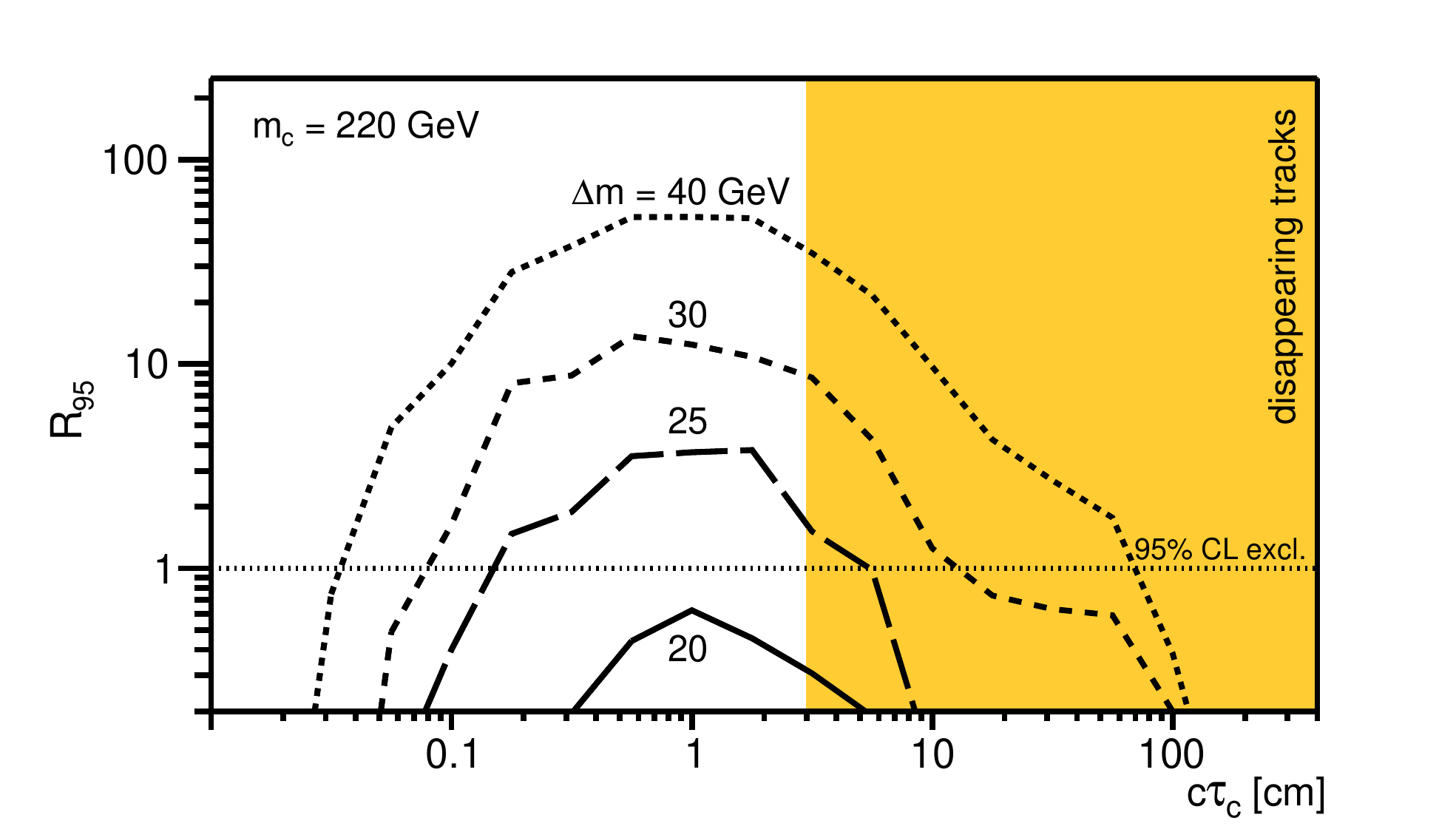}
	\caption{Existing LHC bounds on co-scattering dark matter as a function of the
	%mediator
	decay length of the dark partner, $c\tau_c$.  The concentric curves show the values of $R_{95}$ (defined in Eq.\,\ref{eqn:R95}) from the 8 TeV CMS disappearing track search~\cite{Khachatryan:2014mea} for different compression $\Delta m \sim 20-40 \gev$.  They correspond to pair-production via $pp \rightarrow \chi^+ \chi^-$, followed by $\chi^+ \rightarrow \chi^0\ell \nu$ decays. Maximal bounds expected from the most recent disappearing track search~\cite{Sirunyan:2020pjd} are shown in yellow.
	\label{fig:cms_8tev}}
\end{figure}

In scenarios with compressed dark sectors the leptons are typically much softer than the transverse momentum requirements in previous analyses, as we illustrated in Fig.~\ref{fig:pt-leptons}. If we were to apply the analysis of Ref.~\cite{CMS:2016isf} directly to compressed dark sectors with $\Delta m \lesssim 20-30\gev$, the requirement $p_T(\ell) > 40\gev$ would eliminate a nearly all of events and make the analysis essentially insensitive to compressed sectors.  Even with the 8 TeV search, the situation remains unsatisfactory. We demonstrate the loss of sensitivity of this search at small mass splittings in Fig.~\ref{fig:cms_8tev}.
 %-------------------------------------
%-------------------------------------
 %
  The reach of the search is calculated as a function of the mediator decay length $c\tau_c$ using the log-likelihood ratio $R_{95}$ defined in Eq.~\eqref{eqn:R95}.  As is clearly visible, the sensitivity at $\Delta m = 25\gev$ is already severely reduced, and for $\Delta m = 20\gev$ the search cannot exclude any range of $c\tau_c$ at $95\%$ C.L. A search that targets precisely soft {\it and} displaced kinematics is clearly motivated from this observation.
 
 For the background estimation, we will rely on the $13\tev$ search~\cite{CMS:2016isf}.
  In each of the signal regions from Eq.~\eqref{eq:sr}, background estimates were determined using side bands in data.  The major background in displaced lepton searches are multijet events with heavy flavor jets, which produce $B$ hadrons that decay semi-leptonically at a distance from the production point. Due to statistical fluctuations in the number of hadrons produced during hadronization, as well as reconstruction artefacts, occasionally the lepton from the $B$ decay is misidentified as an isolated displaced lepton.  The estimate is derived in a region with moderate displacements, the so-called displaced control region, and extrapolated to the three signal regions from Eq.~\eqref{eq:sr} using a transfer factor method. 

Using the $\dzero$ distribution in orthogonal $\bbbar$ enriched data, CMS sets 95\% C.L. upper limits to the number of background events in each signal region in 2.6~fb$^{-1}$ of data at $\sqrt{s}=13\gev$.
\begin{align}
\label{eq:cmsyields}
B_{\rm I} = 3.0\,;\qquad B_{\rm II} = 0.50\,;\qquad B_{\rm III} = 0.019\,. 
\end{align}
In each region, multijet events dominated by $\bbbar$ production are the dominant background. All other SM backgrounds, for instance from top-antitop production or electroweak processes, are orders of magnitude smaller.

%%%%%%%%%%%%%%%%%%%%%%%%%%%%%%%%%%%%%%%%%%%%%%%%%
\subsection{Simulation of signal and background events}
\label{sec:simulationsettings}
We describe our signal model using FeynRules~\cite{Alloul:2013bka} and use the UFO~\cite{Degrande:2011ua} interface to {\tt MadGraph5\_aMC@NLO 2.6.6}~\cite{Alwall:2014hca} to generate $2\cdot 10^{6}$ signal events of the process
\begin{align}
pp\to Z^\ast/\gamma^\ast \to \chi^+\chi^- \to (\chi^0 e^\pm\nu)(\chi^0\mu^\mp\nu)
\end{align}
for each of the various benchmarks from Tab.~\ref{tab:benchmarks}. Initial and final state showers as well as hadronization are modelled with {\tt Pythia v8.243}~\cite{Sjostrand:2006za}. As in the CMS analysis~\cite{CMS:2016isf}, we require one electron and one muon of opposite charge in the final state. The cross section $\sigma(m_c)$ for $pp \to \chi^+\chi^-$ is identical to pair production of supersymmetric wino-like charginos. For each benchmark, we rescale our event simulations by the corresponding LHC prediction for $\sqrt{s} = 13\tev$ at NLO+NLL precision~\cite{Fuks:2012qx,Fuks:2013vua}
\begin{align}
   \sigma(220\gev) = 903 \pm 54\,\text{fb} ,\qquad \sigma(324\gev) = 127.7 \pm 9.5\,\text{fb}\,.
\end{align}

 The detector effects for both signal and background samples have been emulated by passing the events through {\tt Delphes 3.4.1}~\cite{deFavereau:2013fsa}, using the CMS detector response.  Electrons and muons are required to fulfill $|\eta(\ell)|<2.4$.  A pre-selection cut of $p_T(\ell) > 15\gev$ is required for all leptons. The lepton isolation criteria are $\textrm{Iso}(e) < 0.12$ and $\textrm{Iso}(\mu) < 0.15$, where isolation is defined as the sum of the $p_T$ of all reconstructed particles within a cone of $R=0.2$ around the lepton, divided by the $p_T$ of the lepton. At analysis level, we set a selection cut of $p_T(\ell) > 20\gev$.  {\tt Delphes} also provides us with information about the lepton transverse impact parameter $\dzero$ and includes a veto for leptons that are created outside the CMS tracking volume.

As the dominant background for our study is from heavy flavor production, the final background will be estimated using the data-driven CMS calculations in Ref.~\cite{CMS:2016isf}, as described in detail in the following section.  However, the shape of the $p_T(\ell)$ spectrum of the background leptons is estimated from Monte Carlo events. To this end we have simulated a large sample of $\bbbar$ events at NLO QCD with {\tt MadGraph5\_aMC@NLO 2.6.6}~\cite{Alwall:2014hca}, followed by hadronization using {\tt Pythia v8.243}~\cite{Sjostrand:2006za}. In order to efficiently generate events with leptonic $B$ decays, we have turned off purely hadronic decay modes of hadrons containing $b$ and $c$ quarks, such that the decay into muons and electrons is dominant. We confirm that the shapes of the $p_T(\ell)$ and $\dzero$ distributions of the leptons remain unchanged for our region of interest. Following this procedure, we generate $2\times 10^7$ di-lepton events via $pp \to \bbbar \to \ell +X$, of which about $3 \times 10^5$ events contain a muon or electron with $p_T(\ell) > 15\gev$. Just like in the analogous method in Ref.~\cite{CMS:2016isf}, lepton isolation is not applied in the determination of the transfer factors and was verified not to affect the $p_T(\ell)$ and $\dzero$ spectra. This procedure improves the fraction of  di-lepton events from $\bbbar$ events by at least three orders of magnitude.

%%%%%%%%%%%%%%%%%%%%%%%%%%%%%%%%%%%%%%%%%%%%%%%%%%%%%%%
\subsection{Background extrapolation to low momenta}\label{sec:background}
We now describe our procedure for estimating the heavy flavor background for relaxed $p_T(\ell)$ requirements.  For the leptons originating from $\bbbar$ decays we observe no large correlation between $p_T(\ell)$ and $\dzero$. A similar feature has been observed in Ref.~\cite{Khachatryan:2014mea}. This allows us to extrapolate the background predictions from Ref.~\cite{CMS:2016isf} to regions with lower $p_T(\ell)$ independently from $\dzero$. The overall normalisation of the background is set by scaling the number of simulated events in the control regions to the event rates in Eq.~\eqref{eq:cmsyields} from Ref.~\cite{CMS:2016isf}.

\begin{figure}[tp]
	\centering
	\includegraphics[width=0.53\textwidth]{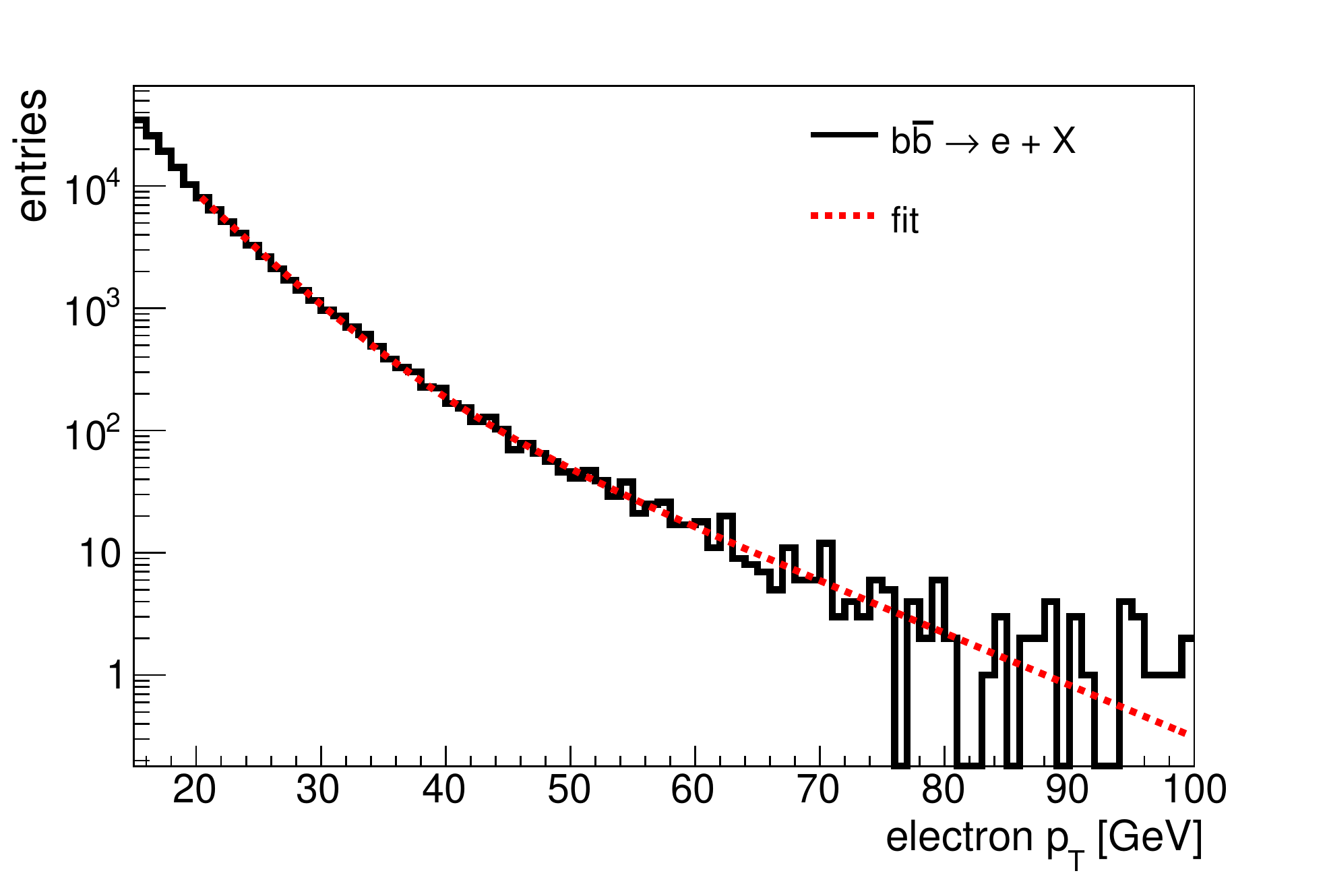}\hspace*{-0.4cm}
	\includegraphics[width=0.53\textwidth]{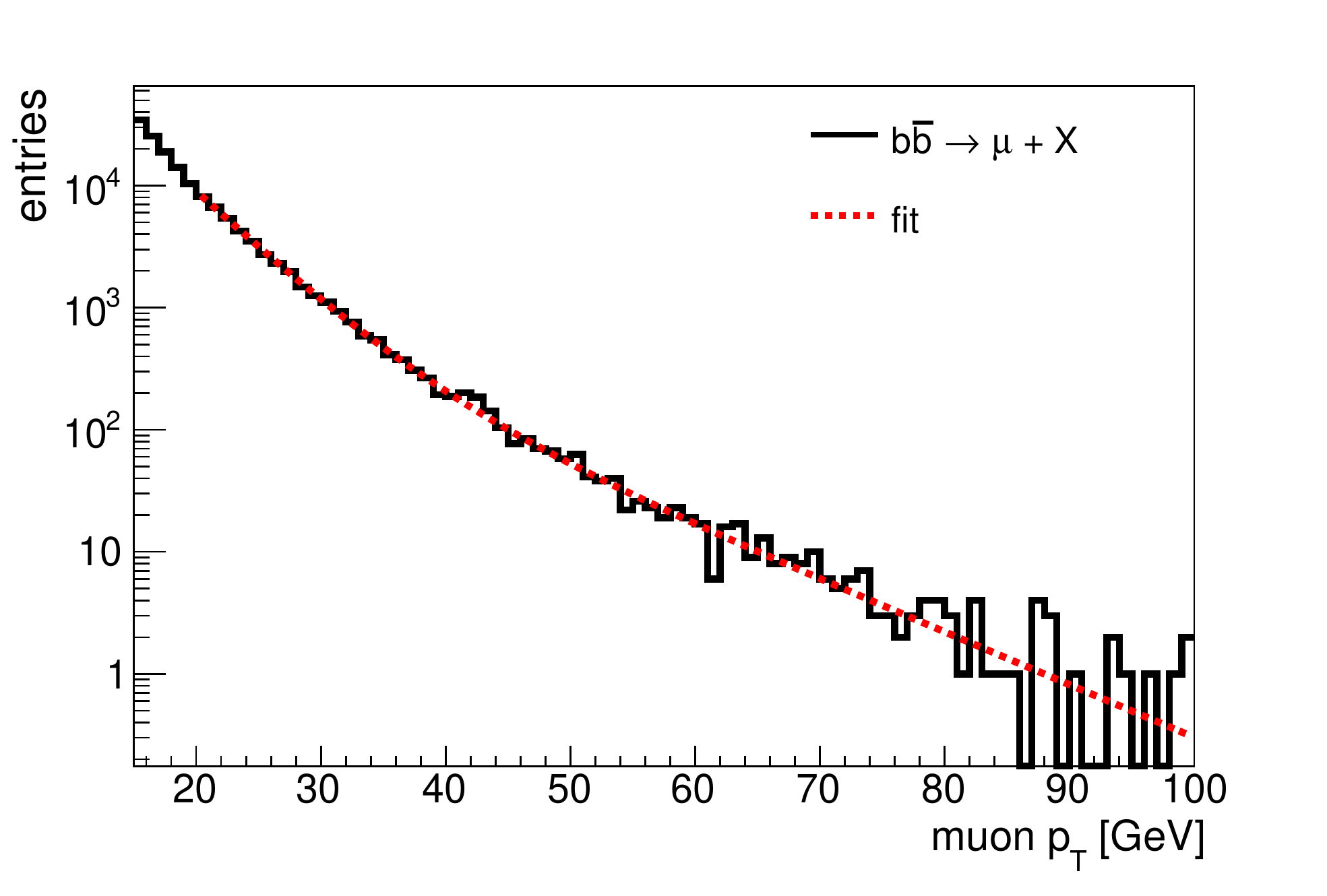}
	\caption{\label{fig:kappafig}The transverse momentum distributions of electrons (left) and muons (right) in $pp\to \bbbar \to \ell + X$ production at $\sqrt{s} = 13\tev$. Shown are our event simulation using {\tt Delphes} (black curve) and the double-exponential fit to the simulation from Eq.~\eqref{eq:transfer} (red dotted curve).}
\end{figure}

In Fig.~\ref{fig:kappafig} we show the transverse momentum distribution of all isolated electrons and muons from HF background processes.
 The distributions are parameterized with a double-exponential fit to allow extrapolation from high to low transverse momenta. We define extrapolation transfer factors for electrons and muons as
\begin{align}\label{eq:transfer}
\kappa_e(p_T) & = \frac{\int_{p_T}^{70} d \tilde{p}_T f_e(\tilde{p}_T)}{\int_{42}^{70} d \tilde{p}_T f_e(\tilde{p}_T)}\,,\qquad f_e(\tilde{p}_T) = e^{13.65 - 0.23 \cdot \tilde{p}_T} + e^{8.62 - 0.10 \cdot \tilde{p}_T}\,,\\\nonumber
\kappa_{\mu}(p_T) & = \frac{\int_{p_T}^{70} d \tilde{p}_T f_\mu(\tilde{p}_T)}{\int_{40}^{70} d \tilde{p}_T f_\mu(\tilde{p}_T)}\,,\qquad f_\mu(\tilde{p}_T) = e^{13.50 - 0.22 \cdot \tilde{p}_T} + e^{8.73 - 0.10 \cdot \tilde{p}_T}\,,
\end{align}
where $p_T$ is given in units of $\gev$. The transfer factors are listed in Tab.~\ref{tab:kappasperlepton} for different thresholds of the lepton momentum $p_T(\ell)$. 
%----------------------------------------------------
\begin{table}[t]
\begin{center}
\renewcommand{\arraystretch}{1.2}
\setlength{\tabcolsep}{4mm}
\begin{small}
\begin{tabular}{c|c|c|c}
\toprule
$p_T(\ell)$ & $\kappa_e(p_T)$ & $\kappa_\mu(p_T)$ & $\kappa_{e\mu}(p_T)$ \\\midrule[1pt]
35 & 2.67$\pm$0.14 & 2.06$\pm$0.12 & 5.51$\pm$0.42  \\ 
30 & 5.94$\pm$0.28 & 4.57$\pm$0.24 & 27.12$\pm$1.92  \\ 
25 & 14.55$\pm$0.66 & 11.04$\pm$0.55 & 160.65$\pm$10.83  \\ 
\textbf{20} & \textbf{ 39.19$\pm$1.73} & \textbf{29.08$\pm$1.45} & \textbf{1139.46$\pm$75.97} \\ \bottomrule[1pt]
\end{tabular}\end{small}
\caption{\label{tab:kappasperlepton}Transfer factors $\kappa_{\ell}(p_T)$ to extrapolate the number of expected background events from Ref.~\cite{CMS:2016isf} to regions with a lower threshold on the lepton transverse momentum $p_T(\ell)$. The selection used in our analysis is indicated in bold font.}
\end{center}
\end{table}
%----------------------------------------------------
%
 As the electrons and muons are decay products of hadrons produced in $\bbbar$ production, their $p_T$ distributions are largely independent from each other. This allows us to define the total transfer factor for the event rate as
\begin{equation}
    \kappa_{e\mu}(p_T)= \kappa_e(p_T) \times \kappa_{\mu}(p_T)\,. \label{eq:kappaformula}
\end{equation}
Using this transfer factor we derive upper limits on the expected background yield $B_i$ in each signal region in terms of the event numbers, $N_{i}$, provided by CMS. The background yield in each signal region $i$ is finally given by
\begin{align}\label{eq:bg}
 B_i = \kappa_{e\mu}(p_T) \times N_{i}
\end{align}
and is listed in Tab.~\ref{tab:eventyields}.

%%%%%%%%%%%%%%%%%%%%%%%%%%%%%%%%%%%%%%%%%%%%%%%%%%%%%%%%%%
\subsection{Event yields for signal and background}\label{sec:yields}
We can now estimate the total number of expected background events from $\bbbar$ production by multiplying the upper limits for different regions in $\dzero$ provided by Ref.~\cite{CMS:2016isf} with the $\kappa_{e\mu}(p_T)$ transfer factors, see Eq.~\eqref{eq:bg}. The resulting background rates in each signal region are listed in Tab.~\ref{tab:eventyields}.
%-----------------------------------------------------------------
\begin{table}[tp]\begin{center}
\renewcommand{\arraystretch}{1.2}
\setlength{\tabcolsep}{4mm}
\begin{small}\begin{tabular}{c|c|c|c}
\toprule
($p_T(e)$,\,$p_T(\mu)$) & $B_{\rm I}$ & $B_{\rm II}$ & $B_{\rm III}$ \\\midrule[1pt]
(42,40) & 3.2 & 0.5 & 0.019 \\  % these are the 95% CL numbers
(35,35)  & 20 & 3.2 & 0.120 \\ % these are the 95% CL numbers
(30,30) & 99 & 15 & 0.587 \\ % these are the 95% CL numbers
(25,25) & 582 & 91 & 3.5 \\ % these are the 95% CL numbers
\textbf{(20,20)} & \textbf{4123} & \textbf{644} & \textbf{25} \\ % these are the 95% CL numbers
\bottomrule[1pt]
\end{tabular}\end{small}
\end{center}
\caption{\label{tab:eventyields}95\% C.L. upper limits on the expected number of HF events in the three signal regions from Eq.~\eqref{eq:sr}, assuming an integrated luminosity of 2.6 fb$^{-1}$ at $\sqrt{s}=13\tev$. The upper limits for (42,40) are adopted from Ref.~\cite{CMS:2016isf}. The selection used in our analysis is indicated in bold font.}
\end{table}
%-----------------------------------------------------------------
%
 As expected, once the $p_T$ threshold for the leptons is lowered, the number of expected background events from $\bbbar$ production increases exponentially.
 
%-----------------------------------------------------------------
\begin{table}[tp]\begin{center}
\renewcommand{\arraystretch}{1.2}
\setlength{\tabcolsep}{4mm}
\begin{small}\begin{tabular}{cl|c|c|c}
\toprule
\text{\#} & ($m_c$\,[GeV], $\Delta m$\,[GeV], $\ctauc$\,[cm]) & $S_{\rm I}$ & $S_{\rm II}$ & $S_{\rm III}$   
\\\midrule[1pt]
1 & (324, 20, 2) & 0.38 & 0.43 & 1.18
\\
2 & (220, 20, 3) & 1.18  & 1.40 & 5.55
\\
3 & (220, 20, 0.1) & 139 & 37 & 5.98
\\
4 & (220, 20, 1) &  174 & 157 & 283 
\\
5 & (220, 20, 10)  &  32 & 93 & 318
\\
6 & (220, 20, 100)  &  1.35 & 2.15 & 31
\\
7 & (220, 40, 1) & 1067  & 980 & 1826 
\\\midrule[1pt]
  &  HF background & 221997 & 34688 & 1318 \\
\bottomrule[1pt]
\end{tabular}\end{small}
\end{center}
\caption{\label{tab:signalyieldsnoNN}Expected number of events for the signal benchmarks and 95\% C.L. upper limit on the HF background events in the three signal regions from Eq.~\eqref{eq:sr}~\cite{CMS:2016isf}, for a threshold of $p_T(\ell) > 20\gev$ on the lepton transverse momentum, assuming an integrated luminosity of 140 fb$^{-1}$ at $\sqrt{s}=13\tev$.}
\end{table}
%-----------------------------------------------------------------

To obtain realistic signal and background yields, the deterioration of the lepton reconstruction efficiency with increasing $\dzero$ must be taken into account.  To model this effect, we make use of the published identification efficiency parametrization from the 8 TeV search~\cite{CMS:2014hka} up to the limit $d_0 = 2$\,cm and linearly extrapolate to 10~cm to cover the full range of the 13 TeV search~\cite{CMS:2016isf}.  After verification that this estimates a realistic signal selection efficiency, the signal cross sections from Sec.~\ref{sec:benchmarks} are used to obtain signal event yields.  

Finally the expected signal and background events for an integrated luminosity of 140~fb$^{-1}$ are listed in Tab.~\ref{tab:signalyieldsnoNN}.  Aside from benchmark 7, which corresponds to the case of large $\Delta m$, none of the benchmarks show a good signal-to-background ratio.  To preserve the sensitivity for the signal when moving the $p_T(\ell)$ threshold to lower momenta, it is therefore essential to further reject the $\bbbar$ background using additional techniques. 

%---------------------------------------------------------------------------
\section{Multi-variate analysis}\label{sec:mva}
 \noindent To assess the expected significance to soft displaced leptons at the LHC during Run 3, we perform a multi-variate analysis using a neural network. In what follows we discuss the topology and kinematic distributions of the leptons in signal and background.  
The use of neural networks is crucial for successfully extracting a signal with soft displaced leptons from LHC data. We show that an analysis with basic kinematic cuts alone does not provide a good signal-background discrimination. However, with a straightforward neural network architecture we achieve a good sensitivity to our signal. This demonstrates a realistic discovery potential of a dark sector through soft displaced leptons at the LHC.

%---------------------------------------------------
\subsection{Kinematic distributions}
\label{sec:variables}
\noindent As discussed in Secs.~\ref{sec:theory} and~\ref{sec:lhcsignal}, in scenarios of co-annihilation and co-scattering leptons from 
%mediator
 decays $\chi^+ \to \chi^{0} \ell \nu$ can be very soft. For $\Delta m = 20 \gev$ their transverse momentum distribution resembles the HF background from Fig.~\ref{fig:kappafig}; for $\Delta m = 40 \gev$ the signal leptons tend to larger $p_T(\ell_1)$ than in the background. For signals with  $\Delta m = 20 \gev$ or lower, that create soft lepton signals, the transverse momentum of the leptons is thus not a good discriminator between signal and background. We therefore systematically explore other kinematic features that characterize the signal and background.

\begin{figure}[tp]
	\centering
	\includegraphics[width=0.52\textwidth]{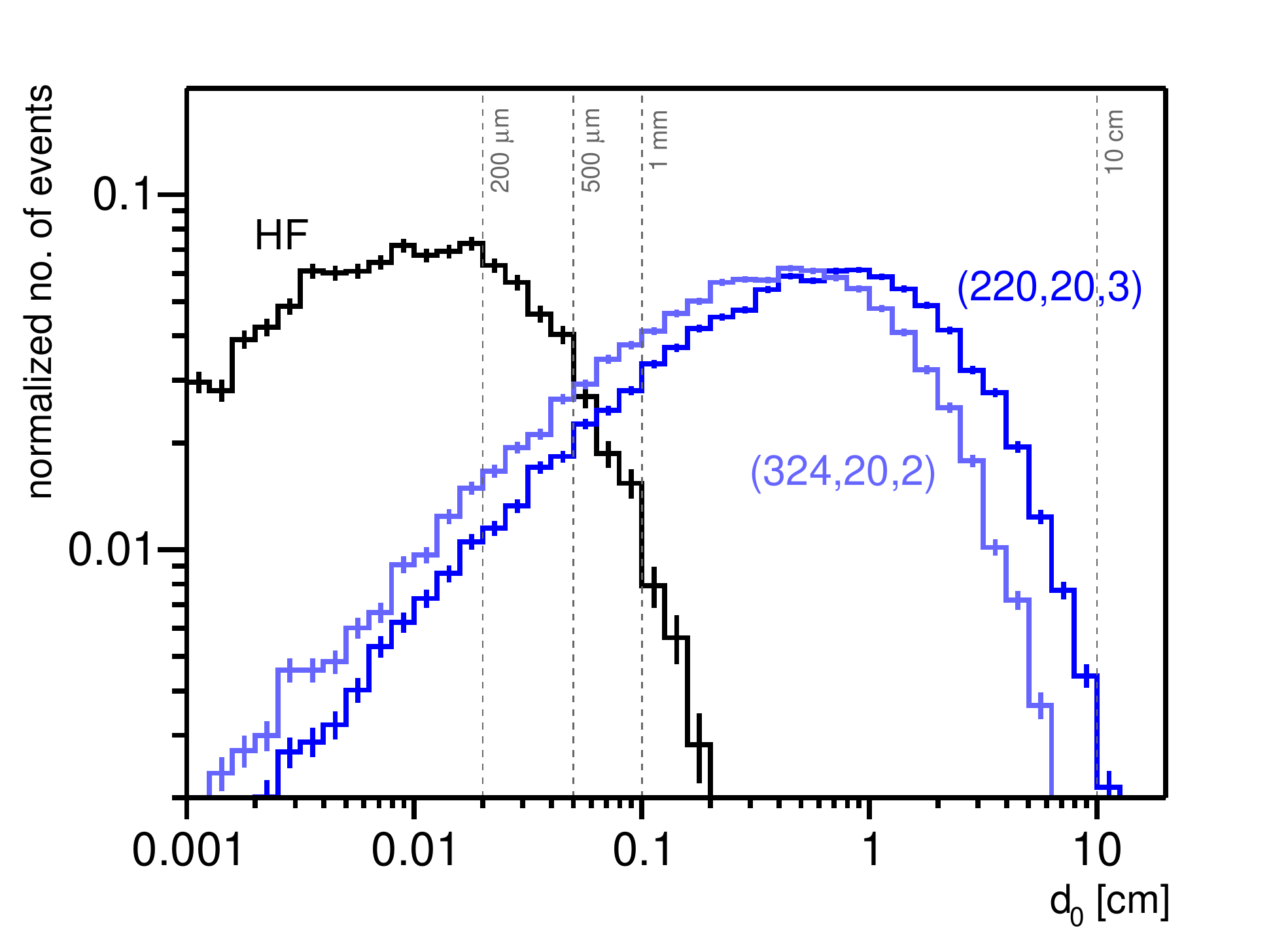}\hspace*{-0.2cm}
	\includegraphics[width=0.52\textwidth]{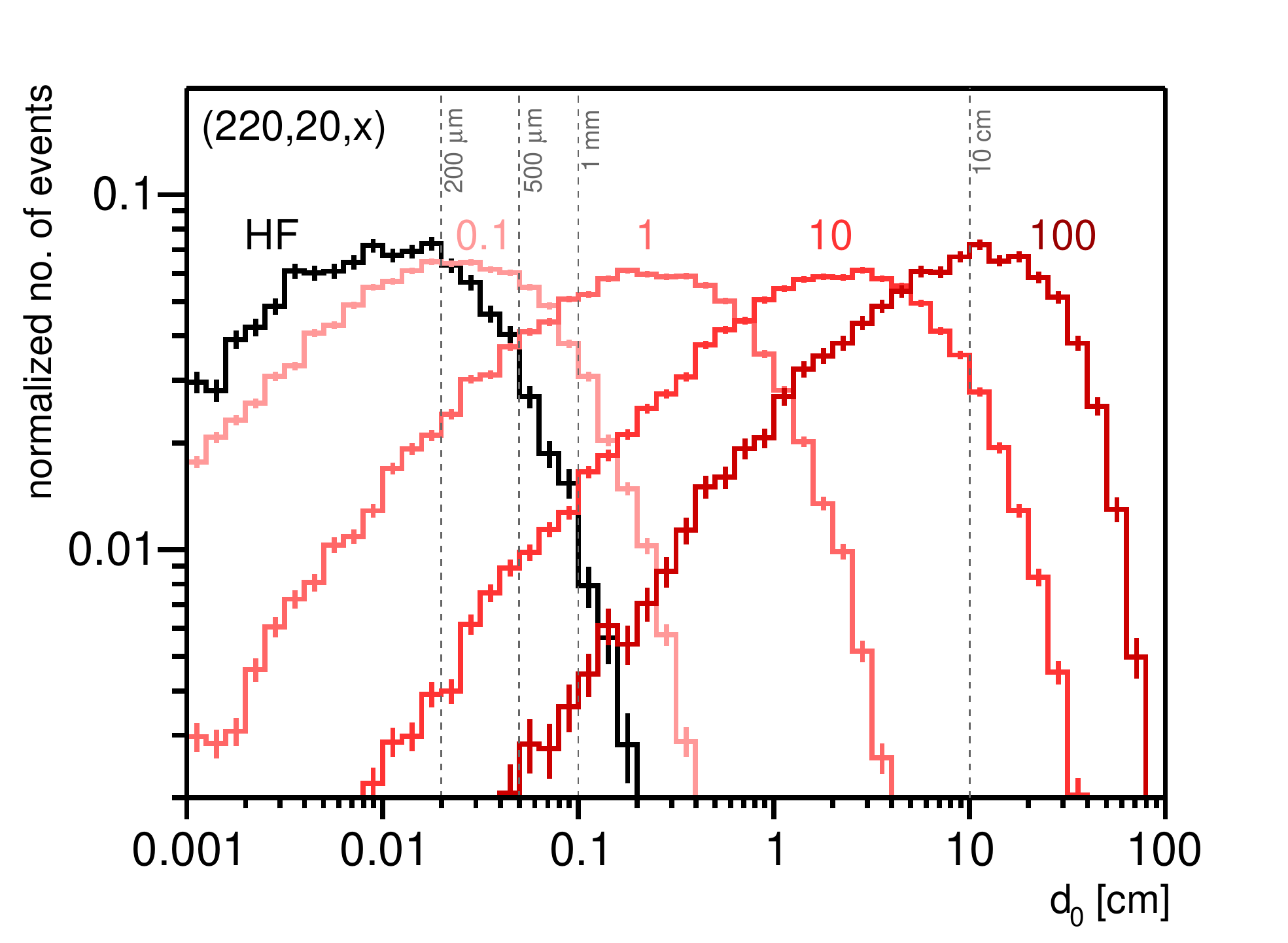}
	\caption{\label{fig:dzero}The transverse impact parameters $\dzero$ of the selected electron-muon pair for various signal benchmarks of dark matter (left) and varying 
	decay length $\ctauc$
	of the charged dark state (right), as well as for the heavy flavor (HF) background. To allow a comparison of the shapes, the distributions are normalized to unity. We do not include $d_0$-based identification efficiencies in this figure, although they are included in the final yields.
	}
\end{figure}

In Sec.~\ref{sec:lhcsignal} we introduced three signal regions distinguished by the transverse impact parameter $\dzero$, which is one of the main discriminants in our analysis. In Fig.~\ref{fig:dzero}, we show normalized distributions of $\dzero$ for the benchmarks defined in Tab.~\ref{tab:benchmarks} and for the HF background. For reference, we indicate the boundaries of the signal regions from Eq.~\eqref{eq:sr} as dashed lines. $\rm SR~I$ is sensitive to 
%mediator
 decay lengths up to $c\tau_c \approx 2\,\text{mm}$, while $\rm SR~II$ and $\rm SR~III$ probe larger decay lengths up to $1\,\text{m}$.
%--------------------------------------
%--------------------------------------
%
 The background peaks around $\dzero = 0.1\,\text{mm}$, as suggested by the decay length of a $B$ meson from the background, $c\tau_B \approx 0.5\,\text{mm}$. For comparison, in benchmark 3:~(220,\,20,\,0.1) (light red), the mediator has a decay length of $c\tau_c = 1\,$mm. All other signal distributions are shifted to larger displacements and allow for a better discrimination from the background.
 
 The impact parameter scales logarithmically with the decay length, reflecting the exponential decay of the dark states,  see Eq.~\eqref{eq:exp-decay}. Therefore the $\dzero$ distributions for benchmarks with exponentially increasing $\ctauc$ are equally separated, as is apparent from Fig.~\ref{fig:dzero}, right. The normalized distributions include the detector acceptance of selecting two leptons that are identifiable in the fiducial tracking volume of the CMS detector. The decrease of detector acceptance at larger decay length is responsible for the shape change that can be observed at impact parameters larger than 50\,cm.
 
\begin{figure}[tp]
	\centering
	\includegraphics[width=0.52\textwidth]{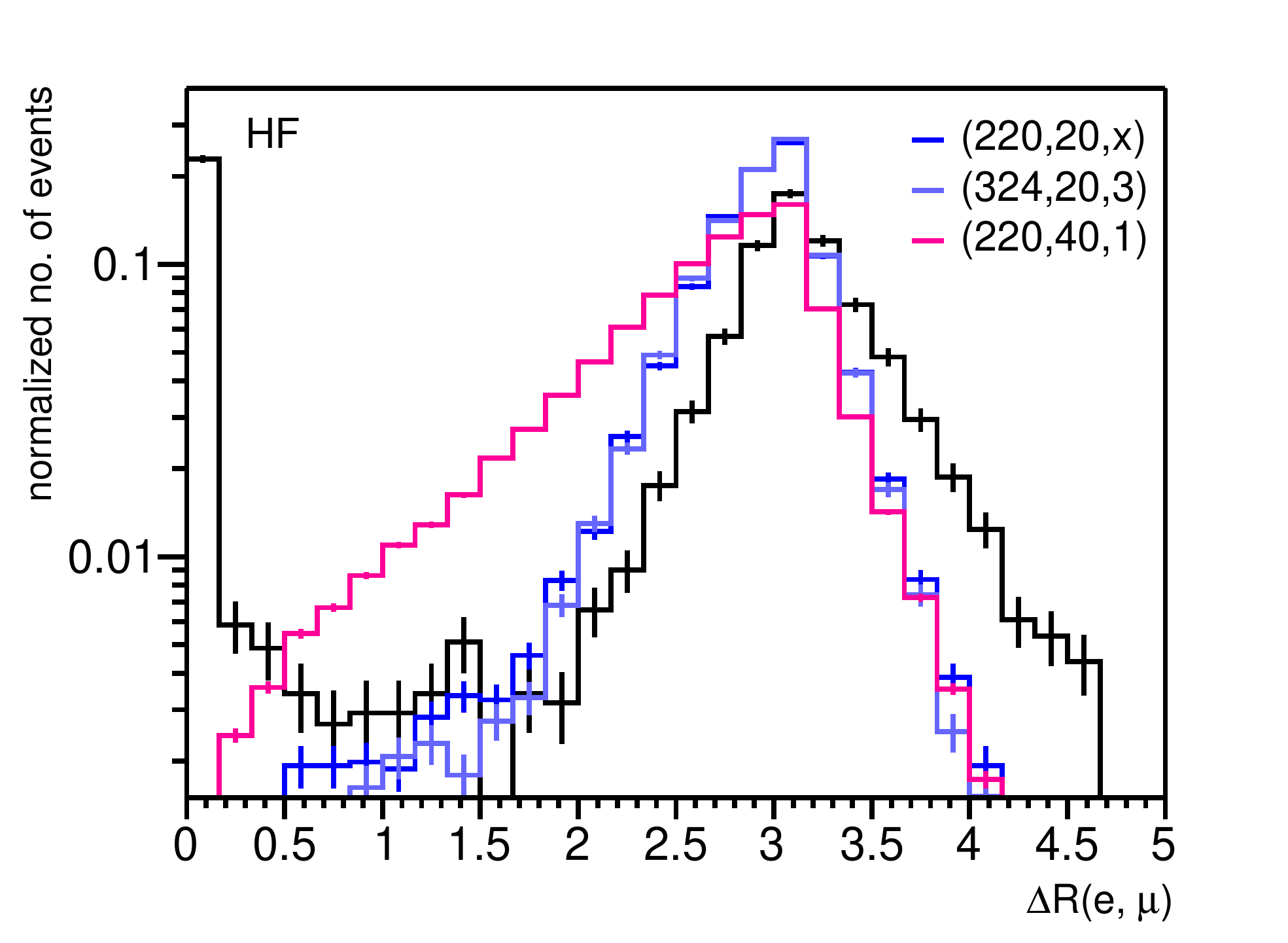}\hspace*{-0.1cm}
	\includegraphics[width=0.52\textwidth]{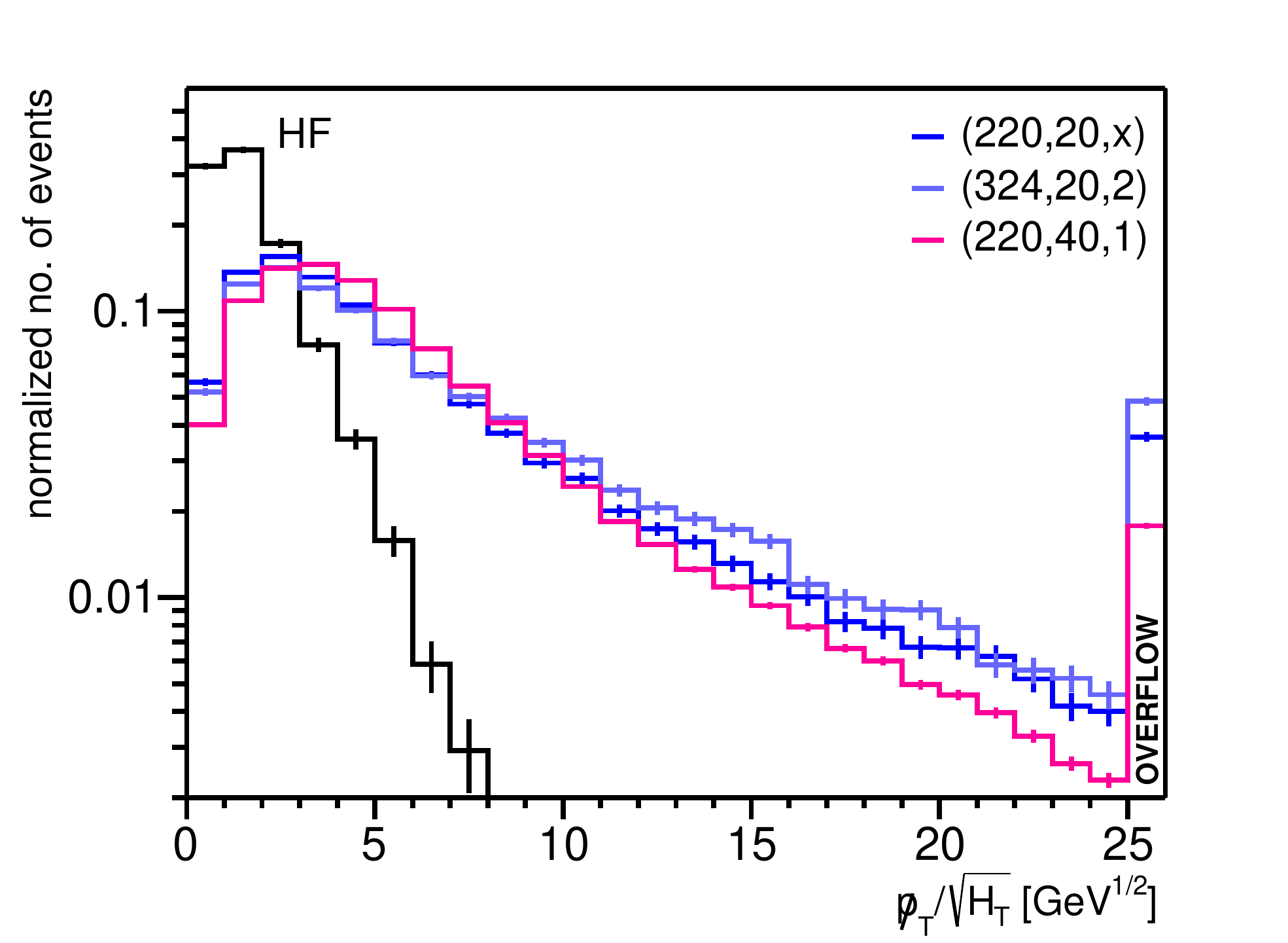}
	\caption{\label{fig:deltaRandMETSignif}Angular lepton separation $\Delta R (e,\mu)$ (left) and missing momentum significance $\met / \sqrt{H_T}$ (right) for representative signal benchmarks and for the heavy flavor (HF) background. To allow a comparison of the shapes, the distributions are normalized to unity.}
\end{figure}

In the center-of-mass frame, $\chi^+$ and $\chi^-$ are produced back-to-back. Due to the overall boost in proton-proton collisions, light particles  are emitted closer to the beamline and with a smaller angular separation. Leptons inheriting the boost of the dark partner are produced with smaller transverse impact parameters $\dzero$ if originating from lighter dark states. This effect is even more pronounced for the very boosted $B$ hadrons from the background. For the benchmarks shown in Fig.~\ref{fig:dzero}, the boost of the dark partner is moderate, resulting in a significant event rate at central pseudorapidity. An important discriminator that explores the back-to-back topology is the angular separation between the signal leptons
\begin{align}
    \Delta R(e,\mu) = \sqrt{\Delta \eta^2(e,\mu)+\Delta \phi^2(e,\mu)}\,.
\end{align}
In Fig.~\ref{fig:deltaRandMETSignif}, left, we show the $\Delta R(e,\mu)$ distributions for three selected signal benchmarks and for the HF background. Soft leptons from decays of heavy dark states tend to be emitted into opposite directions, leading to a maximum around $\Delta R(e,\mu) = \pi$. On the other hand, background leptons are likely to be collimated (anti-collimated) when produced from the same jet (two back-to-back jets). The angular distribution of the background therefore peaks twice and features a different shape than the signal leptons.
%------------------------------------------
%------------------------------------------
 
Additional rejection power can be obtained using combinations of missing momentum, leptons, and any reconstructed jets,\footnote{For the jets we use the default anti-$k_{\tt T}$ definition from {\tt Delphes} and require the selections $p_T(\mathrm{jet}) > 20\gev$, $|\eta(\mathrm{jet})|<2.5$~\cite{Cacciari:2008gp}.} if present. We consider the following kinematic variables: sphericity and spherocity~\cite{Farhi:1977sg,Banfi:2010xy}; $H_T$; the $\met/\sqrt{H_T}$ significance, where both $\met$ and $H_T$ are calculated using only the lepton and jet objects; the transverse mass $m_T(\ell_1,\met)$ calculated with the leading lepton in $p_T$; the azimuthal angle between the leading lepton and the missing transverse momentum, $\Delta \phi(\ell_1,\met)$; and the lepton imbalance $\alpha_T=p_T(\ell_2)/m_T(\ell_1,\ell_2)$~\cite{PhysRevLett.101.221803}, where $\ell_2$ is the sub-leading lepton and $m_T(\ell_1,\ell_2)$ is the transverse mass, calculated from the Lorentz vector sum of the leptons in the transverse plane. 
Each of these variables is known to be robust in a hadron collider environment and verified to be almost completely independent of $\dzero$. We have tested several other kinematic variables commonly used in collider studies, but found them less suitable for signal-background discrimination. We also confirmed that our signal benchmarks for $\Delta m = 20 \gev$ show only modest kinematic differences. 

None of the variables by themselves have enough discriminating power to reject the background sufficiently to compensate for the lower threshold of $p_T(\ell) > 20\gev$. To optimize the sensitivity, we combine them in a multivariate discriminant. Each variable contributes additional sensitivity, while bearing in mind that there are obvious minor correlations between them that, for example, neural nets can use to increase discrimination between the signal and background hypothesis.
%-------------------------------------------------------------
\subsection{Neural network structure}
The nine kinematic variables described in Sec.~\ref{sec:variables} are combined in a neural network. Ranked by performance, these are
\begin{align}
& \{\met/\sqrt{H_T},\ H_T,\ \Delta R (e,\mu),\ m_T(\ell_1,\met),\ (\met/\sqrt{H_T})_{\ell,j},\ \Delta \phi(\ell_1,\met),\\\nonumber
& \hspace*{8cm} \text{sphericity},\, \alpha_T,\, \text{spherocity}\}\,.
\end{align}
To avoid any possible effects of unrealistic modelling of the transverse impact parameter $\dzero$, variables that depend on the displacement or on $\dzero$ itself are not used in the training.

The neural network is implemented in {\sc sci-kit learn} and {\sc tensorflow}, including {\sc keras}~\cite{Pedregosa:2012toh,tensorflow2015-whitepaper}, and includes 2 hidden layers, each with 18 hidden nodes. The hidden layers use activation with a rectified linear unit~\cite{agarap2018deep}, while the output layer uses sigmoid activation so the output probability is summed to unity. For the training procedure, we use the dark matter model with $m_c = 324\gev$, $\Delta m = 20\gev$ and $\ctauc = 2\,$cm, listed as benchmark~1 in Tab.~\ref{tab:benchmarks}, as the signal, together with the HF background sample. The training was performed on 80\% of the HF background sample, and with an equal number of signal events.
As the number of training events is limited by the size of the HF background sample (about 4100 events), dropout regularization is used, and a dropout of 20\% is applied in every hidden layer. The results were verified to be insensitive to over-training, using the remaining 20\% of events that were reserved for independent testing.

%-------------------------------------------------------
\subsection{Performance}
\begin{figure}[tp]
	\centering
	\includegraphics[width=0.52\textwidth]{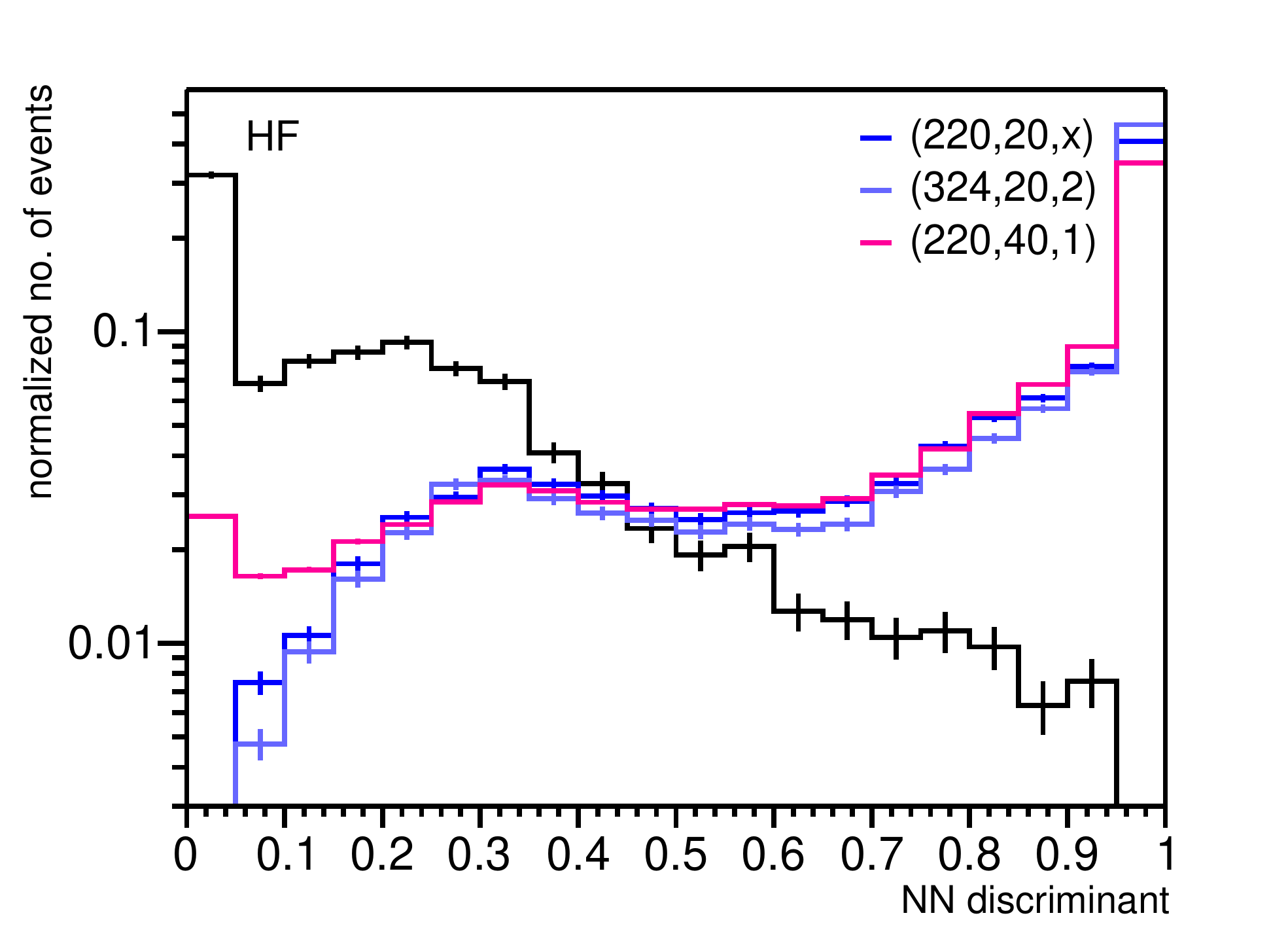}\hspace*{-0.2cm}
	\includegraphics[width=0.52\textwidth]{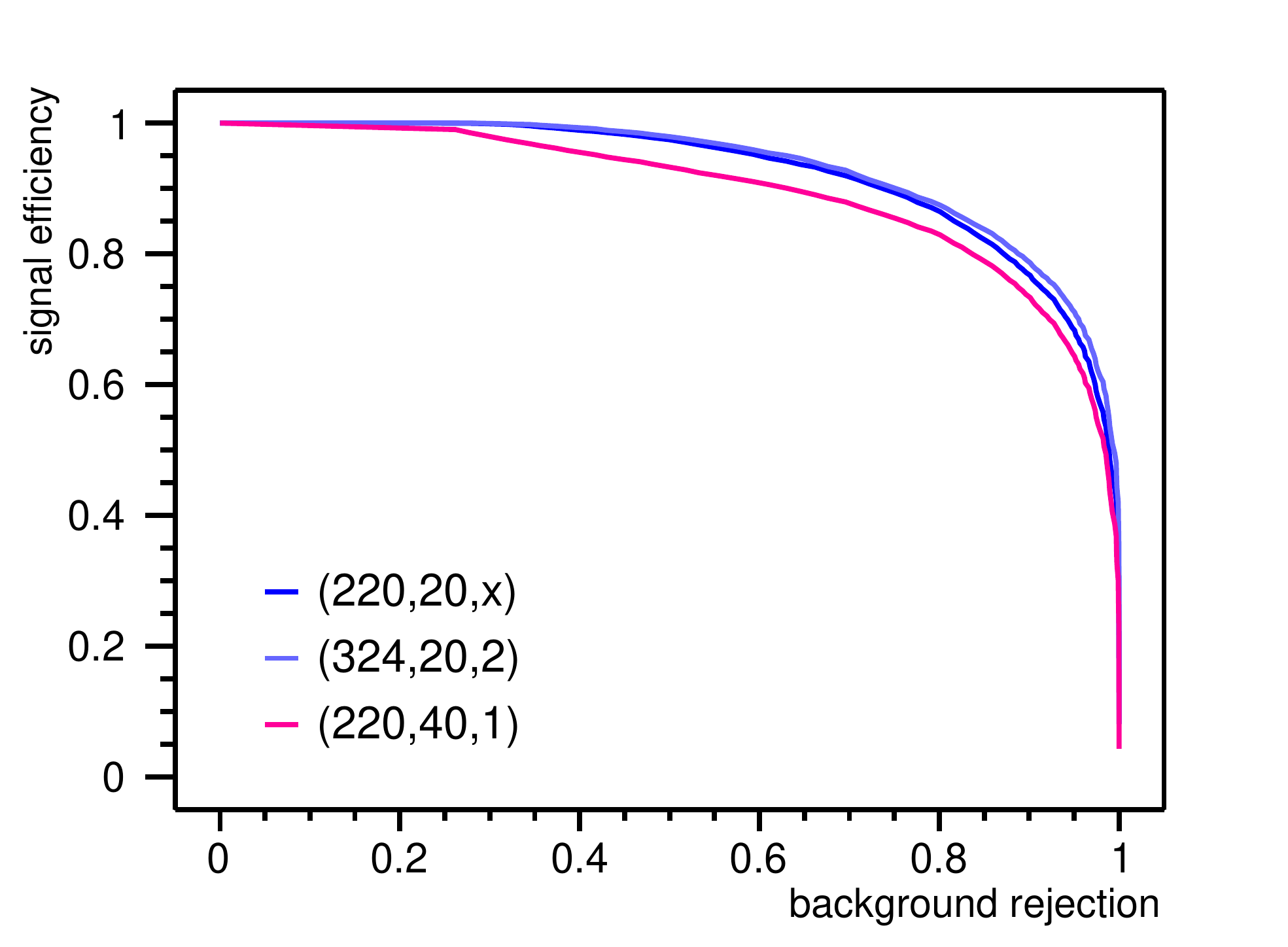}
	\caption{\label{fig:NNfig}The neural network discriminant (left) and ROC curve (right) for selected signal benchmarks and heavy flavor (HF) background. The event yields in the left panel are normalized to unity.}
\end{figure}

The neural network output discriminant, $\nn$, is shown in Fig.~\ref{fig:NNfig}, left, quantifying the discrimination between the HF background and different signal models.
%-------------------------------------------------------
%-------------------------------------------------------
 The output discriminator of the neural network is relatively independent of the benchmark models when the different signal models are compared, as can be observed in the ROC curve in Fig.~\ref{fig:NNfig}, right. This behaviour is expected, because the input variables in Fig.~\ref{fig:deltaRandMETSignif} were chosen to only display minor kinematic differences. The observed similarity suggests a good sensitivity to other benchmarks with soft displaced leptons, without the need to retrain the network.

\begin{figure}[tp]
	\centering
	\includegraphics[width=0.52\textwidth]{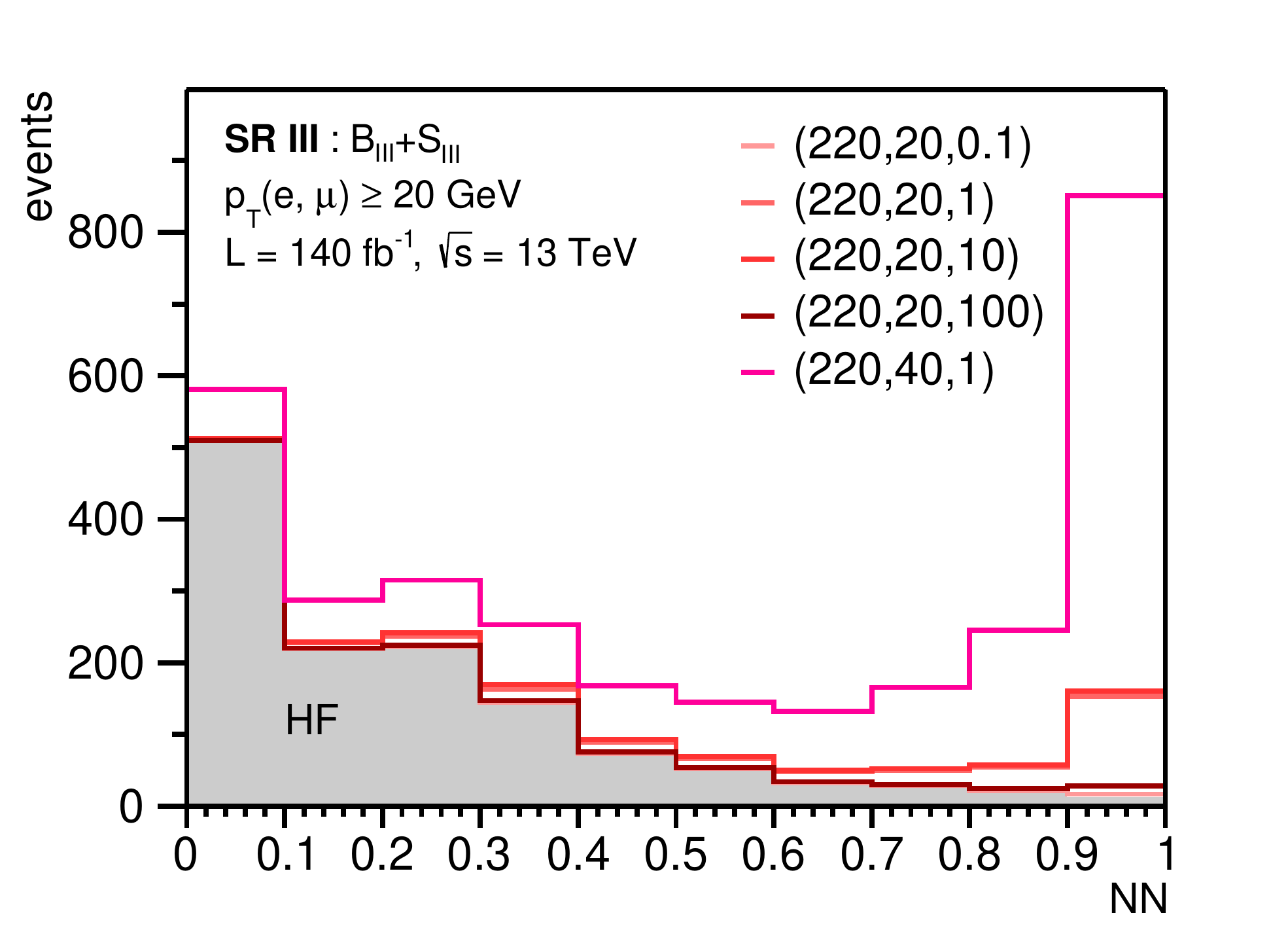}\hspace*{-0.1cm}
	\includegraphics[width=0.52\textwidth]{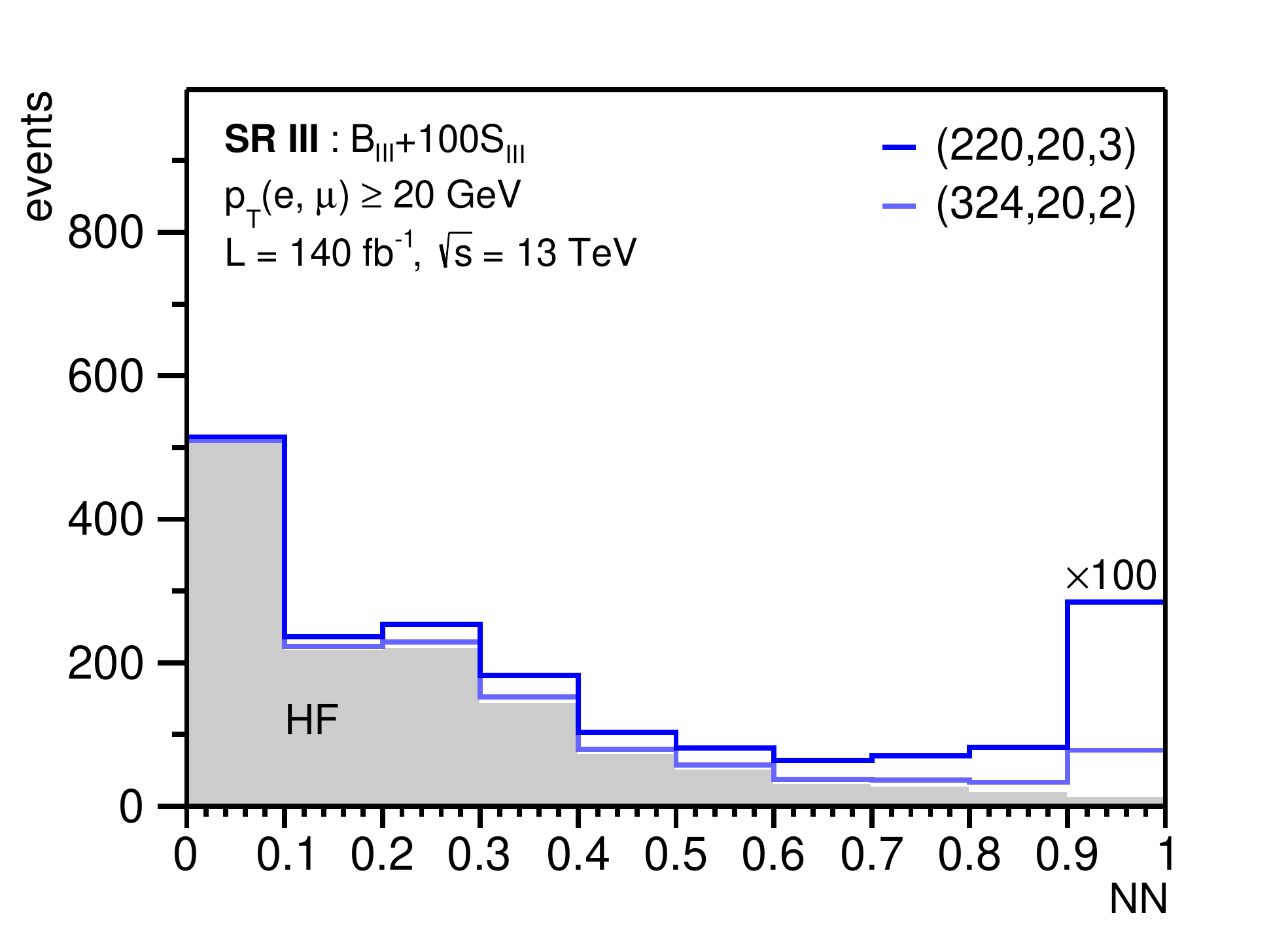}
	\caption{\label{fig:yields}Distributions of the neural network discriminant $\nn$ for the signal benchmarks 3 to 7 (left) and the dark matter benchmarks 1 and 2 (right), and the HF background in the signal region SR III, determined for an integrated luminosity of 140$\,\mathrm{fb}^{-1}$ at $\sqrt{s}=13\tev$. 
	The dark matter benchmarks have relatively low yields. We have scaled the signal distributions by a factor 100 to make them visible to the reader.}
	\end{figure}

 Fig.~\ref{fig:yields} shows the neural network output for an integrated luminosity of 140$\,\mathrm{fb}^{-1}$ in signal region SR III, which has the highest discrimination power between signal and background for most benchmarks. For scenarios with moderate decay lengths, additional sensitivity is gained by combining SR III with the other signal regions. 
 %----------------------------------------------
%----------------------------------------------
The background rejection created by the neural network allows to distinguish between signal and background at high $\nn$ values, thus providing sensitivity to hidden benchmark points. For benchmarks 3 to 6 with leptonically decaying dark states and $\Delta m = 20\gev$ (Fig.~\ref{fig:yields} left), a visible excess at high $\nn$ values can be observed. Applying a hard cut of $\nn \geq 0.9$ reduces the background to less than 1\% of the original value, while retaining a large fraction of the signal. The retained numbers of events are listed in Tab.~\ref{tab:signalyieldswithNN}.
 
For comparison, we also show benchmark 7 with $\Delta m = 40\gev$ (Fig.~\ref{fig:yields} left). The signal-to-background ratio in this scenario is higher than for $\Delta m = 20\gev$ at all $\nn$ values, because the signal leptons generally carry more transverse momentum than those from HF decays. For the dark matter benchmarks 1 and 2 (Fig.~\ref{fig:yields} right) the excess is small, due to the suppressed mediator decays into leptons, see Tab.~\ref{tab:benchmarks}. Further sensitivity could be obtained by performing a binned likelihood fit of the $\nn$ output instead of a cut of $\nn \geq 0.9$. Particularly for the second dark matter benchmark scenario, where over 5 events are expected in the SR III before the cut on the $\nn$, such an approach would have potential.

%-----------------------------------------------------------------
\begin{table}[tp]\begin{center}
\renewcommand{\arraystretch}{1.2}
\setlength{\tabcolsep}{4mm}
\begin{small}\begin{tabular}{cl|c|c|c}
\toprule
\text{\#} & ($m_c$\,[GeV], $\Delta m$\,[GeV], $\ctauc$\,[cm]) & $S_{\rm I}$ & $S_{\rm II}$ & $S_{\rm III}$ 
\\ \midrule[1pt]
1 & (324, 20, 2)   &   0.21  &  0.23 & 0.64 
\\ 
2 & (220, 20, 3)   &  0.57  &  0.67  & 2.71  
\\  
3 & (220, 20, 0.1) &  68   &  19  &  3.06  
\\ 
4 & (220, 20, 1)   &  84   &  72  &  139  
\\ 
5 & (220, 20, 10)  &  15   &  20  &  147  
\\ 
6 & (220, 20, 100) &  0.79   &  0.70  &  14  
\\ 
7 & (220, 40, 1)   &  449   &  427  &  837 
\\ \midrule[1pt]
 & HF background   & 2323   &  363  &  14 
 \\ 
\bottomrule[1pt]
\end{tabular}\end{small}
\end{center}
\caption{\label{tab:signalyieldswithNN}Expected number of events for the signal benchmarks and 95\% C.L. upper limits on the HF background events in the three signal regions from Eq.~\eqref{eq:sr}, for $p_T(\ell) > 20\gev$ and $\nn ~\geq 0.9$ (see text). Assumed is an integrated luminosity of 140 fb$^{-1}$ at $\sqrt{s}=13\tev$.
}
\end{table}
%-----------------------------------------------------------------    

\section{Predictions for the LHC}\label{sec:predictions}
\begin{figure}[tp]
	\centering\hspace*{-0.4cm}
	\includegraphics[width=0.56\textwidth]{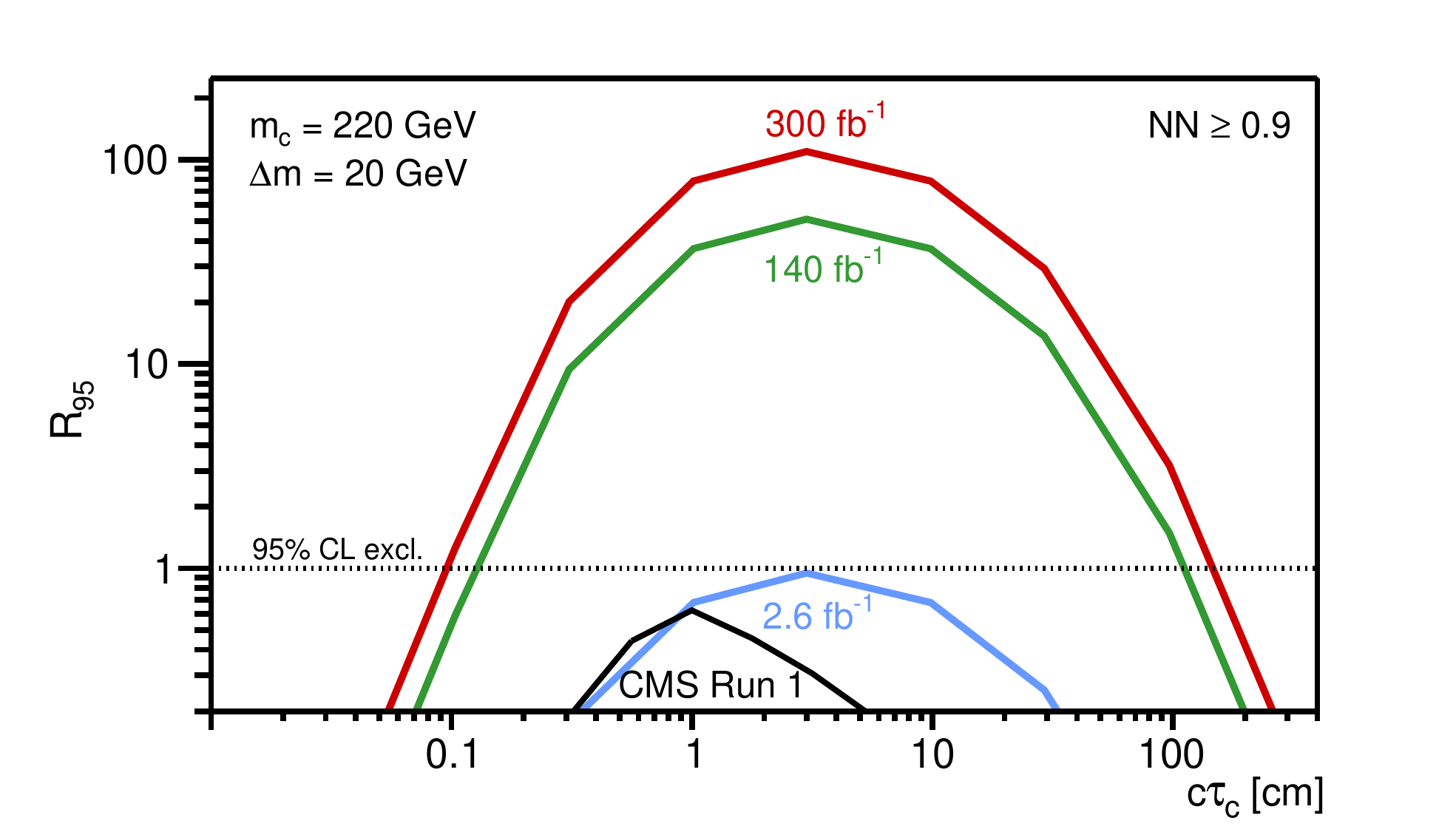}\hspace*{-0.75cm}
	\includegraphics[width=0.56\textwidth]{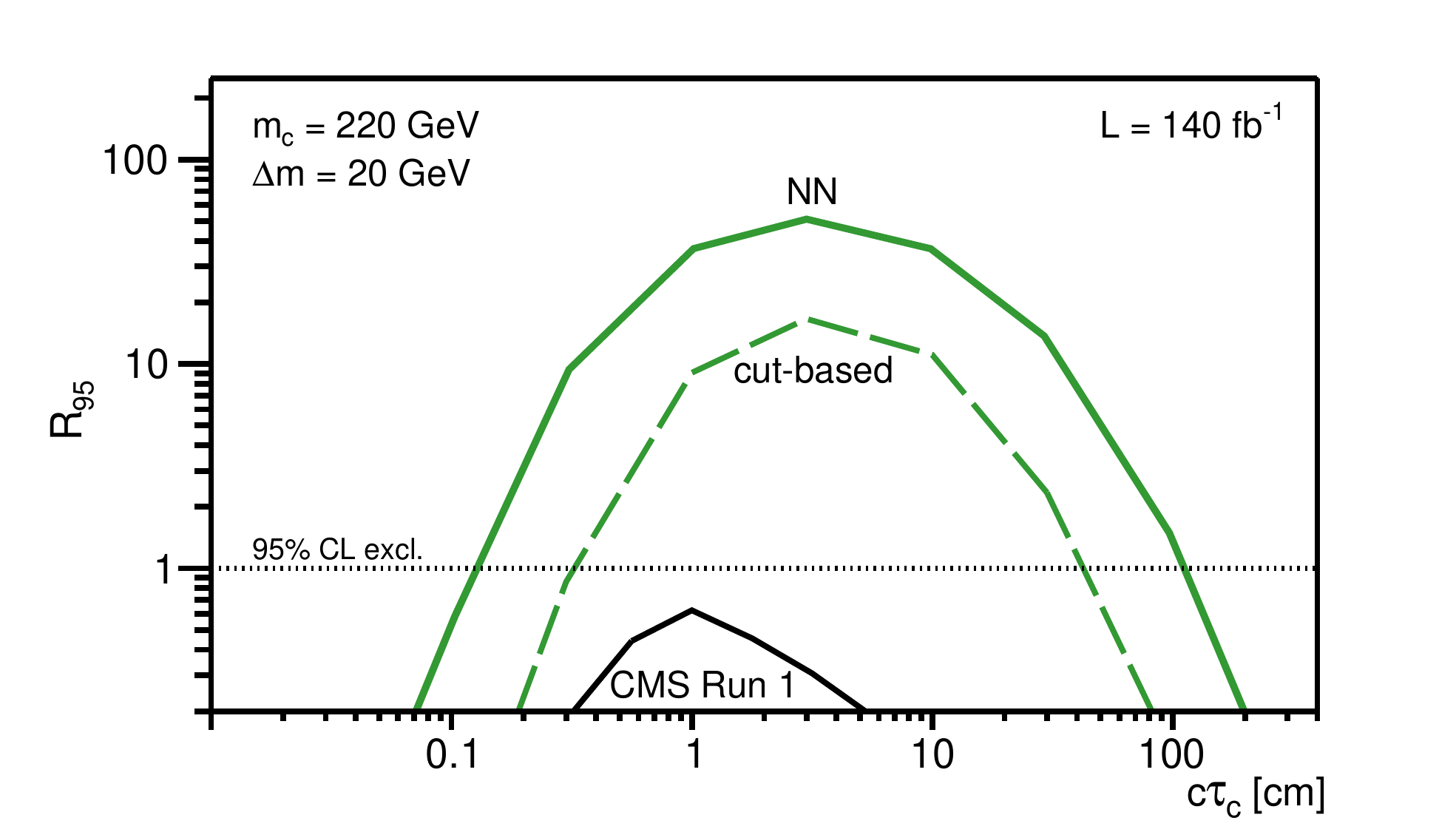}
	\caption{\label{fig:ELplots}Projected signal exclusion limits at the LHC as a function of the	decay length of the charged dark state, $c\tau_c$. The expected sensitivity obtained from a neural network analysis is shown for different integrated luminosities (left) and in comparison with a simple kinematic cut-based analysis for 140~fb$^{-1}$ (right) at $\sqrt{s}=13\tev$. A likelihood ratio $R_{95} > 1$ indicates that a certain decay length is excluded at more than 95\% C.L. For comparison, we show existing upper bounds from a displaced lepton search at $\sqrt{s} = 8\tev$ (plain black curve)~\cite{Khachatryan:2014mea}.}
\end{figure}

\noindent Based on the neural network analysis described in Sec.~\ref{sec:mva}, we investigate the search potential for displaced soft leptons at the LHC. Our results rely on the combined sensitivity in all three signal regions, applying the cut $\nn \geq 0.9$ on the network discriminant in each signal region. In Fig.~\ref{fig:ELplots} we present the expected bounds as a function of the decay length of the charged dark state, illustrating the sensitivity with the data collected during Run 2.

 The presented results are based on individual analyses of the $\nn$ for the soft lepton benchmarks 3-6, which we have interpolated to obtain bounds for intermediate decay lengths. With $140\,\text{fb}^{-1}$ of data, compressed dark sectors with $m_c = 220\gev$ and decay lengths in the range $1\,\text{mm} < c\tau_c < 1\,\text{m}$ can be excluded at the 95\% C.L. This covers essentially the entire range of decay lengths predicted by co-annihilating or co-scattering dark matter, see Sec.~\ref{sec:signal}. The maximum sensitivity is reached at $\ctauc = 3\,\text{cm}$. For benchmarks with smaller or larger $\ctauc$, fewer events fall into the signal region, resulting in a reduced sensitivity at both extremes. The good sensitivity at small decay lengths very close to the detector resolution of $200\,\mu\text{m}$ confirms that the background rejection by our analysis is efficient. The sensitivity at large $c\tau_c$ depends on the experimental efficiency to detect displaced leptons. 
 
In Fig.~\ref{fig:ELplots}, right, we compare the neutral network results with a simple cut-based analysis where only the $p_T(\ell)$ requirement is lowered from 40 to $20\gev$. It is evident that a multi-variate analysis of the event kinematics is crucial to overcome the HF background and obtain a good sensitivity to soft leptons.

We also show the current 95\% C.L. upper bounds from the previous displaced lepton search at $\sqrt{s} = 8\tev$ with $p_T(\ell) > 25\gev$~\cite{Khachatryan:2014mea}, which is not sensitive to displaced soft leptons from dark states produced in weak interactions. The much higher sensitivity of our analysis is due to a combination of the larger data set, the use of a neural network, and the lower lepton threshold. While the $8\tev$ analysis was restricted to $d_0 < 2\,\text{cm}$, the sensitivity in our analysis extends to $\ctauc = 1\,\text{m}$, thus covering the entire tracking region of the CMS detector. This explains why our analysis is sensitive to larger decay lengths, reaching its maximum at $c\tau_c = 3\,\text{cm}$. 

In Fig.~\ref{fig:ELplots}, left, we include our projections for Run 3 (in red), assuming the same detection efficiencies and scaling our event numbers from Run 2. With the higher luminosity of 300$\,\text{fb}^{-1}$, the expected reach can be slightly extended towards both smaller and larger decay lengths. It is also instructive to scale our results down to the data set of 2.6$\,\text{fb}^{-1}$ (in blue), which was used in the previous displaced lepton search~\cite{CMS:2016isf}. We see that even with such a small data set the sensitivity to soft leptons reaches the 95\% C.L. around $\ctauc = 3\,\text{cm}$ and exceeds the very limited reach of the $8\tev$ search.

To reliably reproduce this analysis at the LHC, the data will need to be collected with triggers that have acceptance for leptons with lower transverse momenta. Compared to the triggers used in Ref.~\cite{CMS:2016isf}, it will be essential to lower the $p_T \geq 40\, (42)\gev$ requirement on reconstructed muons (electrons). Flatly decreasing the trigger momentum threshold would result in an unacceptably large increase in the event rate directly proportional to the transfer factor estimated in Tab.\,\ref{tab:kappasperlepton}.  Suitable triggers would place additional requirements on the presence on the angle between the leptons, or can use information from missing transverse momentum or the presence of additional objects such as jets. Recent CMS papers use a missing transverse momentum requirement of approximately 120 GeV that could be suitable and has about a similar background rejection as the neural network. However this trigger is also expected to have tighter thresholds in the future as a function of accelerator performance. To mitigate this increase in $\met$ requirement, a cross trigger that combines displaced lepton information with a $\met$ would be appropriate.  Such cross triggers should allow the rejection of more HF background at trigger level, so that the data collection rates remain at an acceptable level while still being able to collect soft leptons. Another interesting option is to explore the production of dark states from vector boson fusion with dedicated dijet triggers~\cite{Jones-Perez:2019plk}.

%%%%%%%%%%%%%%%%%%%%%%%%%%%%%%%%%%%%%%%%%%%%%%%%%%%%%%%%%
\section{Conclusions and outlook}\label{sec:conclusions}
\noindent We have performed an analysis of soft displaced leptons with $p_T > 20\gev$, produced in association with missing energy at the 13 TeV LHC. This signature is typically predicted from compressed hidden sectors and is particularly well motivated by scenarios of co-annihilation and co-scattering dark matter. Our analysis builds on a previous CMS search for displaced leptons with $p_T > 40\gev$. By lowering the lepton threshold we drastically improve the sensitivity to soft displaced leptons, despite the enhanced background from heavy flavor decays.

Overcoming this background requires a dedicated study of the signal and background kinematics beyond a simple cut-and-count analysis. To this end, we have examined seven benchmarks: two of them motivated by dark matter, four to explore the sensitivity to the displacement for a fixed mediator mass $m_c = 220\gev$ and mass splitting $\Delta m = 20\gev$ in the hidden sector, and one with a larger mass splitting that covers less compressed scenarios. We have trained a neural network with nine selected kinematic variables to maximize the signal sensitivity. For the two dark matter scenarios our search predicts a low sensitivity, because the production rate and the branching ratio of the dark partners into leptons are small for dark partners with electroweak interactions. In turn, we find that dark partners with purely leptonic decays in the range $1\,\text{mm} < \ctauc < 1\,\text{m}$ can be explored with 140 fb$^{-1}$ of current Run 2 data. With 300 fb$^{-1}$ of data collected during Run 3, the sensitivity could be extended to $0.8\,\text{mm} < \ctauc < 2\,\text{m}$. Our search is also sensitive to heavier dark states. For a mass splitting of $\Delta m = 20\gev$, we expect a 95\% C.L. sensitivity to masses up to $500\gev$ with 140 fb$^{-1}$ and $640\gev$ with 300 fb$^{-1}$ of data. Signatures of this kind have so far not been entirely observed in existing searches.  

Beyond the scenarios considered in this work, our search strategy applies generally to models that predict new particles in the $100\gev$ range decaying into soft displaced leptons and missing energy. Specific examples are leptophilic dark matter or models with heavy neutral leptons, which produce the same final state but predict different event kinematics. Optimized searches for some specific scenarios exist but are too model-dependent to be sensitive to co-scattering dark matter and similar models. Our proposed search extends the discovery potential for new physics with displaced leptons beyond existing searches.

To realize our analysis at the LHC, dedicated triggers need to be designed to overcome the enhanced event rate at low lepton momenta. One possibility is the use of cross triggers with additional jets or missing energy together with soft leptons to compensate for the lower lepton threshold. A further possibility may be to use the angular separation between the leptons to reduce the HF background events that pass the trigger. The broad applicability and the promising results of our analysis strengthen the motivation to develop such triggers for ATLAS and CMS.

It would be very interesting to extend the reach of our analysis to longer decay lengths, which would allow us to explore hidden sectors with even smaller couplings. For di-muon signatures, information on the charged track from the inner detector could be combined with the observation of displaced muons in the muon chambers. Combinations of this kind could access decay lengths of a few meters, which are well motivated for instance in freeze-in dark matter scenarios. We leave these ideas for future research and look forward to the first searches for soft displaced leptons at the LHC.

%%%%%%%%%%%%%%%%%%%%%%%%%%%%%%%%%%%%%%%%%%%%%%%%%%%%%%%%%
\begin{center} \textbf{Acknowledgments} \end{center}
\noindent AF acknowledges support of the DFG (German  Research Foundation) through the research training group {\it Particle physics beyond the Standard Model} (GRK 1940). The research of SW has been supported by the Carl Zeiss foundation through an endowed junior professorship and by the German Research Foundation (DFG) under grant no. 396021762--TRR 257. The work by ARS and FB is supported by the Fonds Wetenschappelijk Onderzoek (FWO) under the EOS {\it be.h: The H boson gateway to physics beyond Standard Model} project no.30820817.  ND acknowledges the support of Department of Science and Technology via the Ramanujan Fellowship  SB/S2/RJN-070/2018.

%
%%%%%%%%%%%%%%%%%%%%%%%%%%%%%%%%%%%%%%%%%%%%%%%%%%%%%%%%%

\newpage
\bibliographystyle{JHEP}
\bibliography{main}

%%%%%%%%%%%%%%%%%%%%%%%%%%%%%%%%%%%%%%%%%%%%%%%%%%%%%%%%%%%%%%%%%%%%%%%%%%%

\end{document}